\documentclass[journal, draftclsnofoot,onecolumn,12pt]{IEEEtran}
\usepackage[T1]{fontenc}% optional T1 font encoding

\usepackage{cite}
\ifCLASSINFOpdf
\else
\fi
\usepackage{amsmath}
\usepackage{amssymb }
\usepackage{amsfonts}

\usepackage{graphicx}
\usepackage{lettrine}
\usepackage{epstopdf}
\usepackage{textcomp,booktabs}
\usepackage[usenames,dvipsnames]{color}
\usepackage{colortbl}
\definecolor{mygray}{gray}{.9}
\definecolor{mypink}{rgb}{.99,.91,.95}
\definecolor{mycyan}{cmyk}{.3,0,0,0}
\usepackage{pifont}
\usepackage{subfigure}
\usepackage{multirow}
\usepackage{url}
\usepackage{color}
\usepackage{bm}

\usepackage{algorithm} %format of the algorithm
\usepackage{algorithmic} %format of the algorithm
\usepackage{multirow} %multirow for format of table

\usepackage[dvipsnames,usenames]{color}
\usepackage[usenames,dvipsnames]{color}

\definecolor{light-gray}{gray}{0.90}

\begin{document}

\title{Artificial Intelligence-aided Receiver for A CP-Free OFDM System: Design, Simulation, and Experimental Test}

\author{\IEEEauthorblockN{Jing Zhang$^\ast$, Chao-Kai Wen$^\dagger$, Shi Jin$^\ast$, Geoffrey Ye Li$^\ddag$}\\
\IEEEauthorblockA{$^\ast$ National Mobile Communications Research Laboratory, Southeast University\\
Nanjing 210096, P. R. China, Email: \{jingzhang, jinshi\}@seu.edu.cn\\$^\dag$ Institute of Communications Engineering, Taiwan Sun Yat-sen University\\ Kaohsiung 80424, Taiwan, Email: chaokai.wen@mail.nsysu.edu.tw\\$^\ddag$ School of Electrical and Computer Engineering, Georgia Institute of Technology\\
Atlanta, GA 30332, USA, Email: liye@ece.gatech.edu.}
\thanks{This work was supported in part by the National Science Foundation (NSFC) for Distinguished Young Scholars of China with Grant 61625106, and in part by the NSFC under Grant 61531011. The work of C.-K. Wen was supported by the Ministry of Science and Technology of Taiwan under Grants MOST 107-2221-E-110-026 and the ITRI in Hsinchu, Taiwan.}
}

\maketitle

\begin{abstract}

 Orthogonal frequency division multiplexing (OFDM), usually with sufficient cyclic prefix (CP), has been widely applied in various communication systems. The CP in OFDM consumes additional resource and reduces spectrum and energy efficiency. However, channel estimation and signal detection are very challenging for CP-free OFDM systems. In this paper, we propose a novel artificial intelligence (AI)-aided receiver (AI receiver) for a CP-free OFDM system. The AI receiver includes a channel estimation neural network (CE-NET) and a signal detection neural network based on orthogonal approximate message passing (OAMP), called OAMP-NET. The CE-NET is initialized by the least-square channel estimation algorithm and refined by a linear minimum mean-squared error neural network. The OAMP-NET is established by unfolding the iterative OAMP algorithm and adding several trainable parameters to improve the detection performance. We first investigate their performance under different channel models through extensive simulation and then establish a real transmission system using a 5G rapid prototyping system for an over-the-air (OTA) test. Based on our study, the AI receiver can estimate time-varying channels with a single training phase. It also has great robustness to various imperfections and has better performance than those competitive algorithms, especially for high-order modulation. The OTA test further verifies its feasibility to real environments and indicates its potential for future communications systems.

\end{abstract}

%  keywords

\begin{IEEEkeywords}
OFDM, CP-free, AI, message passing, OTA
\end{IEEEkeywords}

\section{Introduction}

Orthogonal frequency division multiplexing (OFDM) can effectively deal with delay spread of wireless channels and therefore it has been used in almost all wireless systems \cite{2006liye}. To completely mitigate inter-OFDM-block interference (IBI), enough cyclic prefix (CP) must be inserted between adjacent OFDM blocks, which reduces spectral efficiency of OFDM systems, especially when the delay spread is large or the OFDM block duration is short as in many Internet of things (IoT) applications \cite{2015Lorca}. Without sufficient CP, demodulated OFDM signals will suffer from inter-carrier interference (ICI) in addition to IBI.

To deal with the IBI and ICI induced by insufficient CP, several techniques \cite{1998Kim,2004Park,2017SCSE,2015Park,2016MIMO1,2016MIMO2,2017MIMO3} have been proposed for OFDM. An iterative strategy, called residual inter-symbol interference cancellation (RISIC), has been developed in \cite{2004Park,1998Kim,2015Park} to mitigate the {IBI}, which can achieve an acceptable performance if the channel delay spread is moderate. A CP-free OFDM scheme in \cite{2017SCSE}, called symbol cyclic-shift equalization algorithm, includes decision-feedback equalization (DFE) and CP restoration units. To deal with the sensitivity of DFE to the feedback delay, stored feedback equalization is used in \cite{2018SMIC} to eliminate the ISI, which however causes low detection accuracy. The impact of removing the CP in OFDM in massive multiple-input multiple-output (MIMO) systems has been investigated in \cite{2016MIMO1,2016MIMO2,2017MIMO3}. In addition to signal detection, channel estimation is another challenging issue because the received pilot signals are influenced by CP removal. We will use  artificial intelligence (AI) to address both channel estimation and signal detection in a CP-free OFDM system.

%Authors of \cite{2016MIMO1,2016MIMO2,2017MIMO3} considered the impact of removing the CP from OFDM in massive multiple-input multiple-output (MIMO) systems, which assumed that the number of base station antennas tend to approach infinity. Existing algorithms always concentrate on signal detection. Channel estimation is also a huge challenge because the received pilot signals are influenced by CP removal. To address both issues of channel estimation and signal detection, artificial intelligence (AI) should be emphasized.
%Intelligent communication is considered to be one of the mainstream directions in the development of wireless communications after 5G. By introducing artificial intelligence into the wireless communication systems, the performance of the system can be potentially improved. The combination of wireless transmission and deep learning is in the preliminary exploration stage. However, there have been many achievements in wireless physical layer including channel estimation \cite{o2018over}, signal detection \cite{DBLP:journals/corr/abs-1809-09336}, feedback and reconstruction of channel state information \cite{8322184}, channel decoding \cite{8322184, 8482358}, end-to-end wireless communication systems \cite{D2018Deep}.

Intelligent communication is considered one of the mainstream directions in the development of wireless communications after 5G. By introducing AI into wireless communications, system performance can be potentially improved. Deep learning for wireless systems is in the preliminary exploration stage, many achievements have transpired in wireless physical layers \cite{2019liye}, including channel estimation \cite{o2018over}, signal detection \cite{DBLP:journals/corr/abs-1809-09336}, feedback and reconstruction of channel state information (CSI) \cite{8322184, 8482358}, channel decoding \cite{2017decoding}, and end-to-end wireless-communication systems \cite{D2018Deep} \cite{2018globecom}.

Deep learning (DL) has been recently introduced to OFDM receivers owing to its strong ability to perform channel estimation and signal detection, as well as to address the transceiver's imperfection \cite{2017OShea,2017WTQ}. DL approaches can be classified into two categories: data-driven and model-driven \cite{2019liye}. Data-driven DL techniques generally utilize the standard neural network structure as a black box and are trained by a huge data set. In contrast to data-driven DL, model-driven DL methods construct the network topology according to known physical mechanisms and expert knowledge, and therefore require less training data and shorter training time \cite{DBLP:journals/corr/abs-1809-06059}. Both types of DL have been used in OFDM receivers  \cite{OFDM2018DL3,OFDM2018DL2,OFDM2018DL1,8052521liye,gao2018comnet,2018jpw}. A novel data-driven DL architecture has been developed in \cite{OFDM2018DL3} for an OFDM receiver with one-bit complex quantization. In \cite{OFDM2018DL2}, Cascade-Net has been proposed for signal detection in an OFDM system with sufficient CP, where deep neural network is cascaded with a zero-forcing preprocessor to prevent the network stucking in a saddle point or a local minimum point. The Cascade-Net, which can be regarded as model-driven DL, outperforms zero-forcing method and provides robustness against ill conditioned channels. An intelligent OFDM receiver has been proposed in \cite{OFDM2018DL1} based on a deep complex convolutional network. Its performance is comparable to the traditional receiver for a CP-free OFDM system based on expert knowledge in additive white Gaussian noise (AWGN) channels. However, its performance declines precipitously for multi-path fading channel. In \cite{8052521liye}, a fully-connected deep neural network (FC-DNN) has been used for channel estimation and signal detection, which also works well even for a CP-free OFDM system with quadrature phase shift keying (QPSK) modulation. A model-driven DL method, named ComNet, has been developed in \cite{gao2018comnet} to improve the performance of the OFDM receiver, especially with high-order modulation.   Since a recurrent neural network is applied in ComNet, it can improve signal detection performance but has high complexity at the same time. The performance of FC-DNN and ComNet has been compared in \cite{2018jpw}. {The trainable iterative soft thresholding algorithm (TISTA) in \cite{TISTA}, which is also a model-driven method, unfolds the orthogonal approximate message passing (OAMP) algorithm and trains some variables by DL to solve the problem of  sparse signal recovery.  Our previous work, called OAMP detection neural network (OAMP-NET), in \cite{DBLP:journals/corr/abs-1809-09336} and \cite{2019conf}\footnote{{This is the conference version of the paper. Note that the conference paper makes an omission of the citation to TISTA.}}  is inspired by the TISTA. Different from the TISTA, it introduces more trainable parameters and is applied to MIMO detections and CP-free OFDM scenarios.}

%In this article, an AI-aided receiver (AI receiver) for CP-free OFDM system is proposed. The receiver includes channel estimation neural network (CE-NET) and orthogonal approximate message passing (OAMP) detection neural network (OAMP-NET). Compared with that in ComNet \cite{gao2018comnet}, the channel estimation module remains, but the detection structure is completely transformed. The detection part is replaced by an OAMP-NET, which combines the OAMP algorithm and DL by introducing a few trainable parameters. Furthermore, the OAMP-NET has low complexity and is adaptive to different channels and different modulation modes. Simulation results reveal that the proposed OAMP-NET offers remarkable performance and attains lower BER than the existing algorithms with high-order modulation. In addition, a 5G rapid prototyping (RaPro) system is utilized to test the performance for over-the-air (OTA). Testing results also demonstrate the flexibility and robustness of proposed AI receiver.

In this article, we will develop an AI-aided receiver (AI receiver) for a CP-free OFDM system. The receiver includes a channel estimation neural network (CE-NET) and an orthogonal approximate message passing (OAMP) detection neural network (OAMP-NET). Compared with that in ComNet \cite{gao2018comnet}, the channel estimation module is similar, but the detection structure is completely different. In particular, the detection part is replaced by an OAMP-NET, which combines the OAMP algorithm and DL by introducing a few trainable parameters. Furthermore, the OAMP-NET is with low complexity and is adaptive to different channels and modulation modes. As shown by simulation results, the proposed OAMP-NET offers remarkable performance compared with the existing algorithms, especially for a CP-free OFDM system with high-order modulation. Futhermore, with a 5G rapid prototyping (RaPro) system, we perform an over-the-air (OTA) test and demonstrate the flexibility and robustness of the proposed AI receiver.

%The rest of the paper is organized as follows. In section II, system model are presented. Section III introduces the AI receiver. In Section IV, the simulation results are shown. After that, the OTA test is set up. Finally, conclusion is given at Section V.
The rest of the paper is organized as follows. Section II presents the system model. Section III develops the AI receiver. Section IV and Section V provide the simulation results and the OTA test, respectively. Finally, conclusion is given in Section VI.

%Notations: Column vectors are denoted by boldface letters. The superscripts ${{(\cdot )}^{T}}$ and ${{(\cdot )}^{H}}$ represent the transpose and conjugate-transpose, respectively. The Euclidean norm is denoted by $\left\| \cdot  \right\|$. The expectation operator is noted as $\mathbb{E}\{\cdot \}$. In addition, $\mathcal{N}(z:0,{{\sigma }^{2}})$ denotes a real-valued Gaussian random variable $z$ with zero mean and variance ${{\sigma }^{2}}$. Finally, real part and imaginary part of complex are presented by ${\rm Re}\{\cdot \}$ and ${\rm Im}\{\cdot \}$.

Notations: Column vectors are denoted by boldface letters. Superscripts ${{(\cdot )}^{T}}$ and ${{(\cdot )}^{H}}$ represent the transpose and conjugate-transpose, respectively. The Euclidean norm is denoted by $\left\| \cdot  \right\|$. The expectation operator is denoted as $\mathbb{E}\{\cdot \}$. Moreover, $\mathcal{N}(z;0,{{\sigma }^{2}})$ indicates a real-valued Gaussian random variable $z$ with zero mean and variance ${{\sigma }^{2}}$. Finally, the real and the imaginary parts of a complex number are represented by ${\rm Re}\{\cdot \}$ and ${\rm Im}\{\cdot \}$, respectively.

\section{System Model}

%In this section, block diagram of OFDM receiver for CP-free system is presented. Two types of receivers are introduced including model-based OFDM receiver and AI-aided OFDM receiver.
In this section, a block diagram of the receiver for a CP-free OFDM system is presented. Two types of receivers are introduced, including the model-based OFDM and the AI-aided OFDM receivers.
\subsection{CP-Free OFDM System}
%Fig. \ref{fig_1} shows the block diagram of CP-free OFDM system including a transmitter, channel and two types of OFDM receiver. One is model-based CP-free OFDM receiver, the other is AI-aided CP-free OFDM receiver.
%Before the receivers are presented, the transmitter and channel model are introduced first. It is assumed that the $i$th data block is the signal of interest. The input bit $\mathbf{b}_{{}}^{T}=[{{b}_{1}},{{b}_{2}},\ldots ,{{b}_{K}}]$ are modulated as the transmit symbols $\mathbf{u}_{{}}^{T}=[{{u}_{1}},{{u}_{2}},\ldots ,{{u}_{N}}]$ whose duration is ${{T}_{s}}$ and average energy is ${{E}_{s}}$. $K$ and $N$ represents the length of input bits and data symbols, respectively. Assume that transmit symbols are from the same M-ary quadrature amplitude modulation (\emph{M}-QAM) constellation set $\mathcal{A}$ and ${{u}_{n}}={{a}_{n}}+j{{b}_{n}}\in \mathcal{A}$. Then the serial data is conversed to parallel data $\mathbf{u}$ before the inverse FFT (IFFT) block. Next, an $N$-point IFFT is performed on $\mathbf{u}$ to generate an OFDM block $\mathbf{q}$, $\mathbf{q}={{\mathbf{F}}^{H}}\mathbf{u}={{[{{q}_{1}},{{q}_{2}},\ldots ,{{q}_{N}}]}^{T}}$, where $\mathbf{F}$ is a $N\times N$ normalized FFT matrix and is written as

\begin{figure*}[htbp]
		\centering
		\includegraphics[width=7in]{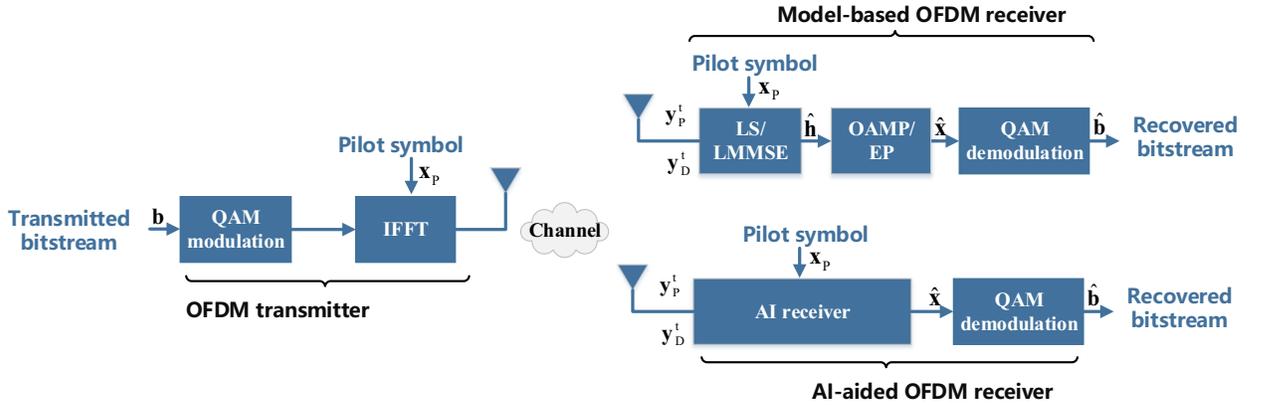}
		% where an .eps filename suffix will be assumed under latex,
		% and a .pdf suffix will be assumed for pdflatex; or what has been declared
		% via \DeclareGraphicsExtensions.
%		\caption{Block diagram of CP-free OFDM system including transmitter, channel and receiver. Model-based receiver utilizes OAMP algorithms while AI receiver integrate DNN into OAMP algorithm.}
        \caption{Block diagram of a CP-free OFDM system including the transmitter, channel, and receiver. The model-based receiver utilizes OAMP algorithms, and the AI receiver integrates DNN into the OAMP algorithm.}
		\label{fig_1}
	\end{figure*}

Fig. \ref{fig_1} shows the block diagram of a CP-free OFDM system, including a transmitter, channel, and the two types of OFDM receivers. At the transmitter, the input bits, $\mathbf{b}_{{}}=[{{b}_{1}},{{b}_{2}},\ldots ,{{b}_{K}}]^{T}$, are modulated into the transmit symbols, $\mathbf{u}_{{}}=[{{u}_{1}},{{u}_{2}},\ldots ,{{u}_{N}}]^{T}$, where $K$ and $N$ represent the lengths of the input bits and the data symbols, respectively. Assume that transmit symbols are from the M-ary quadrature amplitude modulation (QAM) constellation set, $\mathcal{A}$, that is ${{u}_{n}}\in \mathcal{A}$. Then, an $N$-point inverse fast Fourier transform (IFFT) is performed on $\mathbf{u}$ to generate an OFDM signal $\mathbf{q}={{[{{q}_{1}},{{q}_{2}},\ldots ,{{q}_{N}}]}^{T}}$, that is, $\mathbf{q}={{\mathbf{F}}^{H}}\mathbf{u}$, where
%and an average energy of ${{E}_{s}}$
\[\mathbf{F}=\frac{1}{\sqrt{N}}{{\left[ \begin{matrix}
   1 & 1 & \cdots  & 1  \\
   1 & {{W}_{N}} & \cdots  & W_{N}^{(N-1)}  \\
   \vdots  & \vdots  & \ddots  & \vdots   \\
   1 & W_{N}^{(N-1)} & \cdots  & W_{N}^{(N-1)(N-1)}  \\
\end{matrix} \right]}_{N\times N}}\]
%where ${{W}_{N}}\text{=}{{e}^{-j2\pi /N}}$. After that, the parallel data is conversed to serial data ${{\mathbf{q}}^{T}}$ and is transmitted into time-varying wireless channel with additive white Gaussian noise (AWGN) $\mathbf{w}={{[{{w}_{1}},{{w}_{2}},\ldots ,{{w}_{N}}]}^{T}}$, which has independent zero-mean components and $\sigma _{\omega }^{2}$-variance.
%A sample-spaced multipath channel described by complex random variables $\{{{h}_{i}}\}_{l=0}^{I-1}$ is considered \cite{8052521liye}. The delay spread $I-1$, resulting in IBI and ICI, is assumed to be shorter than the length of an OFDM block, namely $I-1<N$.
%
%It should be also noted that the receiver synchronizes with the first path ($i=0$). In addition, the multipath delay ${{\tau }_{i}}$ is assumed as a multiple of ${{T}_{s}}$. In order to obtain channel state information (CSI), the pilot symbols are inserted in the first OFDM block while the transmitted data is appended in the following OFDM blocks. Together they constitute a frame. The channel can be treated as constant spanning over one frame, but change from one to another \cite{8052521liye}.
with ${{W}_{N}}\text{=}{{e}^{-j2\pi /N}}$. After that, ${{{{q}_{1}},{{q}_{2}},\ldots ,{{q}_{N}}}}$ are transmitted into a frequency-selective wireless channel with AWGN, $\mathbf{w}={{[{{w}_{1}},{{w}_{2}},\ldots ,{{w}_{N}}]}^{T}}$, which has independent zero-mean components and variance, $\sigma _{\omega }^{2}$. Different from a typical OFDM system, usually with a CP between adjacent OFDM blocks, we consider a CP-free OFDM system, that is, no CP is inserted between OFDM blocks to maintain high spectral efficiency.
%   s

As in \cite{8052521liye}, we denote the impulse response of channel as $\{{{h}_{i}}\}_{l=0}^{I-1}$, which corresponds a channel length of $I-1$ sample spaces. The channel is assumed to be unchanged during one frame, but change from one to another \cite{8052521liye}. We further assume that the channel length is shorter than that of OFDM blocks, that is $I-1<N$, which is true for almost all OFDM systems. The delay spread of wireless channels will result in IBI and ICI.

To obtain CSI, the first OFDM block consists of pilot symbols. The subsequent OFDM blocks transmit data.
%are inserted in and the transmitted data are appended in the following OFDM blocks. Together, they constitute a frame.

%A sample-spaced multipath channel described by complex random variables $\{{{h}_{i}}\}_{l=0}^{I-1}$ is considered \cite{8052521liye}. The delay spread $I-1$, resulting in IBI and ICI, is assumed to be shorter than the length of an OFDM block, namely $I-1<N$. Note also that the receiver synchronizes with the first path ($i=0$). In addition, the multipath delay ${{\tau }_{i}}$ is assumed as a multiple of ${{T}_{s}}$. To obtain CSI, the pilot symbols are inserted in the first OFDM block and the transmitted data are appended in the following OFDM blocks. Together, they constitute a frame.
%The model-based CP-free OFDM receiver contains LS/LMMSE channel estimation module, OAMP signal detection module and QAM demodulation module. Channel estimation module estimates CSI in frequency domain first. Then the estimated CSI is transformed into time domain by performing IFFT. The OAMP module recover the modulated symbol utilizing the estimated CSI in time domain. Finally, the bitstream are recovered by QAM demodulation module.

For a CP-free OFDM system with $N$ subcarriers, the received signal vector $\mathbf{y}=[{{y}_{1}},{{y}_{2}},\ldots ,{{y}_{N}}]^T$ can be expressed as
\begin{align}
   \mathbf{y}
  &=\mathbf{Cq}-\mathbf{Aq}+\mathbf{A}{{\mathbf{q}}_{-1}}+\mathbf{w} \\\nonumber
 &=\mathbf{C}{{\mathbf{F}}^{H}}\mathbf{u}-\mathbf{A}{{\mathbf{F}}^{H}}\mathbf{u}+\mathbf{A}{{\mathbf{q}}_{-1}}+\mathbf{w},
\end{align}
%where $\mathbf{w}$ is the additive white Gaussian noise vector, $\mathbf{u}$ represents the transmit symbol vectors, ${\mathbf{q}}$ and ${\mathbf{q}}_{-1}$ represent the current and the previous OFDM signal vectors, and $\mathbf{F}$ is a $N\times N$ normalized FFT matrix,
where ${\mathbf{q}}$ and ${\mathbf{q}}_{-1}$ represent the current and the previous OFDM signal vectors, and
\[\mathbf{C}={{\left[ \begin{matrix}
   {{h}_{0}} & 0 & \cdots  & 0 & {{h}_{I-1}} & \cdots  & {{h}_{2}} & {{h}_{1}}  \\
   {{h}_{1}} & {{h}_{0}} & 0 & \cdots  & 0 & {{h}_{I-1}} & \cdots  & {{h}_{2}}  \\
   \vdots  & \ddots  & \text{ } & \text{ } & \text{ } & \ddots  & \text{ } & \vdots   \\
   0 & \cdots  & 0 & {{h}_{I-1}} & {{h}_{I-2}} & \cdots  & {{h}_{1}} & {{h}_{0}}  \\
\end{matrix} \right]}}\]
is an $N\times N$ cyclic channel matrix corresponding to the current OFDM signal vector,
and
\[\mathbf{A}={{\left[ \begin{matrix}
   0 & \cdots  & 0 & {{h}_{I-1}} & \cdots  & \cdots  & {{h}_{1}}  \\
   0 & \cdots  & 0 & 0 & {{h}_{I-1}} & \cdots  & {{h}_{2}}  \\
   \vdots  & \cdots  & \vdots  & \ddots  & \ddots  & \ddots  & \vdots   \\
   0 & \cdots  & 0 & \ddots  & \ddots  & 0 & {{h}_{I-1}}  \\
   \vdots  & \cdots  & \vdots  & \ddots  & \ddots  & \ddots  & \vdots   \\
   0 & \cdots  & 0 & 0 & \ddots  & \cdots  & 0  \\
\end{matrix} \right]}}\]
is an $N\times N$ cut-off channel matrix corresponding to the previous OFDM signal vector.

The second and third terms of (1) represent ICI and IBI, respectively. If there exists a sufficient CP in the OFDM system, then $\mathbf{A=0}$ and no ICI or IBI is present.

The received signal in (1) can be further transformed into
\begin{align}
\nonumber
   \mathbf{y}&=\left( \mathbf{C}-\mathbf{A} \right){{\mathbf{F}}^{H}}\mathbf{u}+\mathbf{A}{{\mathbf{q}}_{-1}}+\mathbf{w} \\
 & ={{\mathbf{J}}}{{\mathbf{F}}^{H}}\mathbf{u}+\mathbf{A}{{\mathbf{q}}_{-1}}+\mathbf{w}
\end{align}
where
\[{{\mathbf{J}}}={{\left[ \begin{matrix}
   h_0 & 0 & \cdots  & \cdots  & \cdots  & 0  \\
   \vdots  & \ddots  & \ddots  & \text{ } & \text{ } & \vdots   \\
   h_{I-1} & \text{ } & \ddots  & \ddots  & \text{ } & \vdots   \\
   0 & \ddots  & \text{ } & \ddots  & \ddots  & \vdots   \\
   \vdots  & \ddots  & \ddots  & \text{ } & \ddots  & 0  \\
   0 & \cdots  & 0 & h_{I-1}  & \cdots  & h_0  \\
\end{matrix} \right]}}.\]
%Let $\mathbf{s}={{\mathbf{J}}}{{\mathbf{F}}^{H}}\mathbf{u}$ denote the signal received in the time domain. Thus, the signal-to-noise ratio at the receiver can be expressed as
%\begin{equation}\label{snr}
%  \text{SNR}=10{{\log }_{10}}({{\bar{s}}}/{\sigma _{w }^{2}}),
%\end{equation}
%where ${\bar{s}}=\mathbf{E}\{ {{\left| \mathbf{s} \right|}^{2}} \}$ and $\sigma _{w}^{2}$ represents the variance of $\mathbf{w}$.
is a $N\times N$ matrix.

Let $\mathbf{s}={{\mathbf{J}}}{{\mathbf{F}}^{H}}\mathbf{u}$ denote the signal received in the time domain. Thus, the signal-to-noise ratio at the receiver can be expressed as
\begin{equation}\label{snr}
  \text{SNR}=10{{\log }_{10}}({{E_s}}/{\sigma _{w }^{2}}),
\end{equation}
where ${E_s}=\mathbf{E}\{ {{\left| \mathbf{s} \right|}^{2}} \}$ and $\sigma _{w}^{2}$ represents the variance of $\mathbf{w}$.

\subsection{Model-based Receiver}
As in Fig. \ref{fig_1}, the model-based CP-free OFDM receiver contains a least-squares/linear minimum mean-squared error (LS/LMMSE) channel estimation module, an OAMP signal detection module and a QAM demodulation module. The channel estimation module estimates the CSI in the frequency domain first. Then, the estimated CSI is transformed into the time domain by performing IFFT. The OAMP module recovers the modulated symbol by utilizing the estimated CSI in time domain. Finally, the transmit bits are recovered by the QAM demodulation module.

\subsection{AI Receiver}

%As Fig. \ref{fig_1} shown, AI receiver module and QAM demodulation module constitute AI-aided CP-free OFDM receiver. Compared with model-based CP-free OFDM receiver, it introduce AI into the channel estimation module and signal detection module, respectively. The AI receiver module is detailed in Fig. \ref{figreceiver} including CE-NET and OAMP-NET. The CE-NET is developed from channel eatimation subnet in ComNet \cite{gao2018comnet} due to its predictability and accuracy. The input of CE-NET, $\mathbf{y}_{p}$, is the received pilot signal. It first is input a serial-to-parallel (S/P) converter and then is performed FFT. The initialization block, LS, is equipped with least-square (LS) channel estimation to get ${{\mathbf{\hat{H}}}_{\text{LS}}}$ which initializes subsequent neural network to generate accurate channel estimation, $\mathbf{\hat{H}}$. The signal detection is implemented by OAMP-NET which is a model-driven DNN combining OAMP algorithm and DNN. In particular, the OAMP-NET uses OAMP algorithm in \cite{2017OAMP} as the initialization and then adds some adjustable parameters to improve the detection performance.

\begin{figure}[!h]
\centering
\includegraphics[width=4.5in]{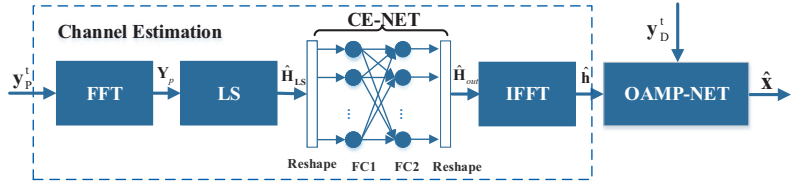}
% where an .eps filename suffix will be assumed under latex,
% and a .pdf suffix will be assumed for pdflatex; or what has been declared
% via \DeclareGraphicsExtensions.
\caption{Block diagram of advanced receiver for a CP-free OFDM system.}
\label{figreceiver}
\end{figure}

 As in Fig. \ref{fig_1}, the AI receiver module and QAM demodulation module constitute the AI-aided CP-free OFDM receiver. Compared with the model-based CP-free OFDM receiver, the AI-aided counterpart introduces AI into the channel estimation module and the signal detection module. The AI receiver module is shown in Fig. \ref{figreceiver}, including channel estimation and an OAMP-NET.

 The channel estimation is developed from the channel estimation subnet as in ComNet \cite{gao2018comnet} because of its predictability and accuracy. The input of channel estimation, $\mathbf{y}_{\rm {p}}^{\rm {t}}$, is first converted into the frequency domain by FFT. Then the block, LS, performs the least-squares (LS) channel estimation to obtain ${{\mathbf{\hat{H}}}_{\text{LS}}}$, which initializes the subsequent neural network to generate more accurate channel estimation, $\mathbf{\hat{H}}_{\rm out}$.

 Signal detection is implemented by the OAMP-NET, which is a model-driven DNN combining the OAMP algorithm and DNN. In particular, the OAMP-NET uses the OAMP algorithm in \cite{2017OAMP} as the initialization and then adds some adjustable parameters to improve the detection performance.

\section{AI Receiver}
%In this section, the AI receiver is utilized to solve the signal detection problem for CP-free OFDM system, which consists of CE-NET and OAMP-NET. CE-NET is based on LS and LMMSE channel estimation which adds the principle of neural network. The OAMP-Net is, developing from the OAMP algorithm, a model-driven deep learning network which can further improve the performance of OAMP algorithm for CP-free OFDM system.
In this section, the AI receiver is developed to solve the signal detection problem for a CP-free OFDM system, which consists of the channel estimation module and the OAMP-NET. The CE-NET is based on LS and LMMSE channel estimation as in \cite{gao2018comnet}. Different from the OAMP algorithm, the OAMP-Net is a model-driven DL network, which can further improve the performance for a CP-free OFDM system.
\subsection{CE-NET}
The LS channel estimation, ${{\mathbf{\hat{H}}}_{\text{LS}}}$ at each subcarrier $n$ , is obtained by
\[{{\mathbf{\hat{H}}}_{\text{LS}}(n)}={\mathbf{Y}_{{\rm {p}}}(n)}/{\mathbf{X}_{{\rm {p}}}(n)},\]
where $\mathbf{Y}_{{\rm {p}}}(n)$ and $\mathbf{X}_{{\rm {p}}}(n)$ represent the received pilot signal and transmit pilot symbol in frequency domain, respectively.

As in \cite{cho2010mimo}, the LMMSE channel estimation can be obtained by
\begin{equation}\label{lmmse}
 {{\mathbf{\hat{H}}}_{\text{LMMSE}}}={{\mathbf{W}}_{\text{LMMSE}}}{{\mathbf{\hat{H}}}_{\text{LS}}}={{\mathbf{R}}_{\mathbf{H\hat{H}_{\text{LS}}}}}{{\left( {{\mathbf{R}}_{\mathbf{HH}}}+\frac{\sigma _{\omega}^{2}}{E_s}\mathbf{I} \right)}^{-1}}\mathbf{\hat{H}_{\text{LS}}},
\end{equation}
where ${\mathbf{W}}_{\text{LMMSE}}$ denotes the $N\times N$ complex weight matrix. The corresponding real-valued form can be expressed as
\begin{equation}\label{6}
  {{\mathbf{\tilde{H}}}_{\text{LMMSE}}}={{\mathbf{\tilde{W}}}_{\text{LMMSE}}}{{\mathbf{\tilde{H}}}_{\text{LS}}},
\end{equation}
where
\[{{\mathbf{\tilde{H}}}_{\text{LMMSE}}}=\left[ \begin{matrix}
 {\rm Re}\{{{{\mathbf{\hat{H}}}}_{\text{LMMSE}}}\} \\
  {\rm Im}\{{{{\mathbf{\hat{H}}}}_{\text{LMMSE}}}\}
\end{matrix} \right],
{{\mathbf{\tilde{H}}}_{\text{LS}}}=\left[ \begin{matrix}
 {\rm Re}\{{{{\mathbf{\hat{H}}}}_{\text{LS}}}\} \\
 {\rm Im}\{{{{\mathbf{\hat{H}}}}_{\text{LS}}}\} \\
\end{matrix} \right],\]
and
\[{{\mathbf{\tilde{W}}}_{\text{LMMSE}}}=\left[ \begin{matrix}
    {\rm Re}\{{{\mathbf{W}}_{\text{LMMSE}}}\} & {\rm Im}\{{{\mathbf{W}}_{\text{LMMSE}}}\}  \\
   {\rm Im}\{{{\mathbf{W}}_{\text{LMMSE}}}\} &  {\rm Re}\{{{\mathbf{W}}_{\text{LMMSE}}}\}  \\
\end{matrix} \right].\]

%The components for CE-NET in Fig. \ref{figreceiver} are FFT, LS, CE-NET and IFFT.
%Then ${{\hat{\bf H}}_{{\rm{LS}}}}$ is used by the CE-NET to generate accurate ${\hat{\bf H}}$ through an one-layer DNN with specially designed weights initializations. Specifically, the complex-valued LS channel estimation  is firstly reshaped into a real-valued vector, where the real part and the imaginary part are concatenated, since the DNNs are basically based on real-valued operations. Next, the real-valued LS vector is as the input of fully connected (FC) layer, in which the multiplicative parameters are initialized by the real-valued linear minimum mean-squared error (LMMSE) weight matrix ${{\tilde{\bf W}}_{{\rm{LMMSE}}}}$ in the (4). Assumed that the additive parameters are initialized by zeros and the activation function is not taken into consideration, the output of the first forward propagation of the CE-NET is the result of LMMSE channel estimation, which can be promoted by the subsequent optimization of the DNN training process.

 As shown in Fig. \ref{figreceiver}, the channel estimation module consists of FFT, LS, CE-NET, and IFFT. Then, ${{\hat{\bf H}}_{{\rm{LS}}}}$ is used by the CE-NET to generate more accurate ${\hat{\bf H}}_{\rm out}$ through an one-layer DNN with specially designed initial weights. The CE-NET is a simple neural network with one input layer and one output layer. The weights are initialized by using the real-valued LMMSE channel estimation weight, ${{\mathbf{\tilde{W}}}_{\text{LMMSE}}}$. The biases are initially set as zero. The input of the CE-NET is real-value  ${{\mathbf{\tilde{H}}}_{\text{LS}}}$, which has $2N$ neutrons. The number of neurons in the output layer is also $2N$. These output neurons have no activation function. The CE-NET is trained by minimizing the the $\ell_2$ loss between predictions and known channel samples by using a specific optimizer. The weights and bias are updated by back propagation algorithm \cite{2008Haykin} during the training process.
\subsection{OAMP-NET}
%Recently, iterative detectors based on approximate message passing (AMP) have been proposed for MIMO detection \cite{DBLP:journals/corr/abs-1809-09336,2017OAMP,2014AMP}, which have low complexity and are easy to be implemented in practice. Therefore, we apply the OAMP algorithm to signal detection for a CP-free OFDM system.
%The interference from the previous OFDM blocks, corresponding to the second term in (2), must be eliminated to apply the OAMP algorithm. As depicted in Fig. \ref{figreceiver}, the receiver acquires the CSI, $\mathbf{\hat{h}}$, by CE-NET. After removing residual ISI, the received signal can be expressed as

Recently, iterative detectors based on approximate message passing (AMP) have been proposed for MIMO detection \cite{DBLP:journals/corr/abs-1809-09336,2017OAMP,2014AMP} since they have low complexity and are easily implemented in practice. Therefore, we will apply the OAMP algorithm for signal detection in a CP-free OFDM system. The interference from the previous OFDM blocks, corresponding to the second term in (2), must be eliminated to apply the OAMP algorithm. As depicted in Fig. \ref{figreceiver}, the receiver acquires the CSI, $\mathbf{\hat{h}}$, by CE-NET. After removing the residual IBI, the received signal can be expressed as
\begin{align}
   \mathbf{\hat{y}}&=\mathbf{y}-\mathbf{{A}}{{{\mathbf{\hat{q}}}}_{-1}} \\\nonumber
 & ={{\mathbf{J}}}{{\mathbf{F}}^{H}}\mathbf{u}+\mathbf{A}{{\mathbf{q}}_{-1}}-\mathbf{{A}}{{{\mathbf{\hat{q}}}}_{-1}}+\mathbf{w} \\\nonumber
 & = \mathbf{{H}u}+\mathbf{w'},
\end{align}
where $\mathbf{{J}}$ and ${\mathbf{A}}$ are derived by the estimated $\mathbf{\hat{h}}$, $\mathbf{{H}}={{\mathbf{{J}}}}{{\mathbf{F}}^{H}}$, and $\mathbf{w'}=\mathbf{A}{{\mathbf{q}}_{-1}}-\mathbf{\hat{A}}{{{\mathbf{{q}}}}_{-1}}+\mathbf{w}$ with a variance of $\sigma _{\omega ' }^{2}$. The OAMP-NET is then performed to detect the transmitted OFDM block.

%Let $\mathbf{s}={{\mathbf{H}}}\mathbf{u}$ denote the signal received in the time domain. Thus, the signal-to-noise ratio at the receiver can be expressed as
%\begin{equation}\label{snr}
%  \text{SNR}=10{{\log }_{10}}({{\bar{s}}}/{\sigma _{w' }^{2}}),
%\end{equation}
%where ${\bar{s}}=\mathbf{E}\{ {{\left| \mathbf{s} \right|}^{2}} \}$.

As the OAMP-NET only addresses real-valued variables, the complex-valued OFDM system is converted into the corresponding real-valued one before OAMP detection as (5) in Section III.A. {To simplify the symbolic notations, new real-version variances are omitted and the complex-version variances represent the real ones below. For example, the real version of $\mathbf{{H}}$ can be noted as $\tilde {\mathbf{{H}}}$, but we continue use $\mathbf{{H}}$ to represent $\tilde {\mathbf{{H}}}$.}

%%算法1
\begin{algorithm} %算法开始
\caption{OAMP Algorithm for CP-free OFDM} %算法的题目
\label{alg1} %算法的标签
\begin{algorithmic}[1] %此处的[1]控制一下算法中的每句前面都有标号
\REQUIRE
$\mathbf{y}$: the received signal in time domain; \\
\hspace*{0.21in}SNR: Signal to Noise Ratio (dB);
%输入条件(此处的REQUIRE默认关键字为Require，在上面已自定义为Input)
\ENSURE
$\mathbf{\hat{b}}$: recovered signal\\
%输出结果(此处的ENSURE默认关键字为Ensure在上面已自定义为Output)
\STATE
\textbf{initialize}: $L=10$, $\beta=0.5$, $\upsilon _{0}^{2}=0$, ${\mathbf{\hat{u}}}_{1}=0$
\STATE acquire the $\hat{\mathbf{h}}$ in real time by the trained CE-NET \\
\STATE compute $\mathbf{{J}}, \mathbf{{A}}$ according to $\hat{\mathbf{h}}$\\
\STATE obtain the noise power by (3)\\
\STATE attain $\mathbf{\hat{y}}$, $\mathbf{{H}}$, $\mathbf{{w'}}$ by (6)\\
%\WHILE {(1)}
\FOR{$l=1$ to $L$}
  \STATE
  execute OAMP algorithm:
    \begin{align}
      & {{\mathbf{r}}_{l}}={{\mathbf{\hat{u}}}_{l}}+{{\mathbf{P}}_{l}}(\mathbf{\hat y}-\mathbf{H}{{\mathbf{\hat{u}}}_{l}}) \\
     & \upsilon _{l}^{2}=\frac{{\left\| \mathbf{\hat y}-\mathbf{H}{{\mathbf{\hat{u}}}_{l}} \right\|}^2-M\sigma _{\omega '}^{2}}{\rm tr({{\mathbf{H}}^{H}}\mathbf{H})} \\
     &\tilde{\upsilon} _{l}^{2}=(1-\beta)\upsilon _{l-1}^{2}+\beta \upsilon _{l}^{2}\\
     & {{\tau }_{l}^{2}}=\frac{1}{2N}\mathrm {tr}({{\mathbf{B}}_{l}}\mathbf{B}_{l}^{H})\tilde{\upsilon} _{l}^{2}+\frac{1}{4N}tr({{\mathbf{P}}_{l}}\mathbf{P}_{l}^{H})\sigma _{\omega '}^{2}\\\label{ul}
      & {{{\mathbf{{\hat{u}}}}}_{l+1}}=\mathbb{E}\{\mathbf{u}|{{\mathbf{r}}_{l}},{{\tau }_{l}}\}
    \end{align}
    \STATE $l=l+1$
\ENDFOR
%\STATE compute $\mathbf{\hat{u}}=\mathbf{\hat{u}}_{L+1}$
\STATE demodulate $\mathbf{\hat{u}}_{L+1}$ to obtain $\mathbf{\hat{b}}$.
\end{algorithmic}
\end{algorithm}
The OAMP-based detector \cite{2017OAMP} can be summarized as Algorithm 1.
In (7), $l$ indicates the index of the iteration times. According to \cite{DBLP:journals/corr/abs-1809-09336}, the optimal matrix ${{\mathbf{P}}_{l}}$ is given by
\begin{equation}\label{w}
  {{\mathbf{P}}_{l}}=\frac{2N}{\mathrm{tr}({{{\mathbf{\hat{P}}}}_{l}}\mathbf{H})}{{\mathbf{\hat{P}}}_{l}}
\end{equation}
where ${{\mathbf{\hat{P}}}_{l}}$ is the LMMSE matrix,
\begin{equation}\label{w_hat}
  {{\mathbf{\hat{P}}}_{l}}=\upsilon _{l}^{2}{{\mathbf{H}}^{H}}{{(\upsilon _{l}^{2}\mathbf{H}{{\mathbf{H}}^{H}}+\frac{\sigma _{\omega ' }^{2}}{2}\mathbf{I})}^{-1}}
\end{equation}
The matrix ${{\mathbf{B}}_{l}}$ in the algorithm is given by ${{\mathbf{B}}_{l}}=\mathbf{I}-{{\mathbf{P}}_{l}}\mathbf{H}$.

Updating the OAMP solution is critical to its stability, particularly for high-order modulations. We first review the parameter updating methods used in the related approaches \cite{2018EPJ,2011EPM,2014EP2,DBLP:journals/corr/abs-1809-09336,2018EPturbo}, and then explain the ones used in Algorithm 1. Following \cite{2014EP2}, some parameters must be tuned, including the minimum allowed variance, $\epsilon$, the damping procedure, $\beta$, and the number of iterations, $L$. The first parameter guarantees non-negativity and the second one determines the stability and the speed of convergence of the algorithm. The computational complexity of the algorithm relates linearly with the third one, $L$. As the EP algorithm in \cite{2014EP2}, the updating of parameter , $\upsilon _{l}^{2}$, can be smoothened using a damping parameter with the former value as shown in (9) to improve the stability of the proposed algorithm. The damping parameter, $\beta$ is set to be 0.2 and 0.95 in \cite{2014EP2} and \cite{2018EPJ}, respectively, $\beta \text{=}\min ({{\exp }^{t/1.5}}/10,0.7)$ in \cite{2018EPturbo}. However, no damping parameters are present in \cite{2011EPM, DBLP:journals/corr/abs-1809-09336, 2017OAMP}. In Algorithm 1, the damping update parameter, $\beta$, is set to be 0.5, which is tuned manually according to various simulation results. The calculation result in (10) should be non-negative. Thus, $\tau_{l}^{2}$ is replaced by $\max{(\tau _{l}^{2},\epsilon)}$ for a small positive constant, $\epsilon=1.0\times10^{-9}$. In all the methods above, the tuning process of the damping parameters is hand-crafted and thus is with low efficiency. Here, we propose a novel structure, called OAMP-NET, to render the parameter setting more flexibly and effectively.

\begin{figure}[!h]
\centering
\includegraphics[width=4in]{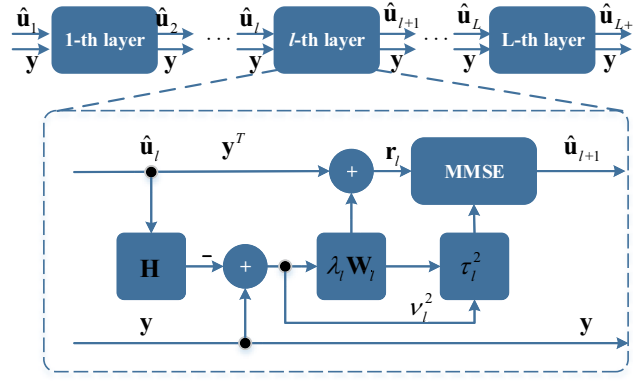}
\caption{Structure of the OAMP-NET.}
\label{figOAMPNet}
\end{figure}

From (\ref{ul}), ${{\mathbf{r}}_{l}}$ and $\tau _{l}^{2}$ are the prior mean and variance, respectively, which influence the accuracy of ${{\mathbf{\hat{u}}}_{l+1}}$. We use the OAMP-NET to provide an appropriate step size to update ${{\mathbf{r}}_{l}}$ and $\tau _{l}^{2}$ and learn the optimal variables from labeled data.
The structure of the OAMP-NET is illustrated in Fig. \ref{figOAMPNet}. The network consists of $L$ cascade layers, each with the same structure that contains the MMSE denoiser, error mean ${{\mathbf{r}}_{l}}$, error variance $\tau _{l}^{2}$, and trainable weights. The input of the OAMP-NET includes the received signal, $\mathbf{y}$, and the initial value, ${{\mathbf{\hat{u}}}_{1}}\text{=}0$. The output is the estimated signal symbol, ${{\mathbf{\hat{u}}}_{L+1}}$. For the \emph{l}-th layer of the OAMP-NET, the input includes the estimated signal, ${{\mathbf{\hat{u}}}_{l-1}}$, from the \emph{l}-1-th layer and the received signal, $\mathbf{y}$.

The OAMP-NET introduces two scalar trainable parameters, $(\lambda _{l}^{{}},\gamma _{l}^{{}})$, which differs from the OAMP algorithm. Thus, the OAMP algorithm is transformed into
\begin{align}
&{{\mathbf{r}}_{l}}={{{\mathbf{u}}}_{l}}+{{\lambda }_{l}}{{\mathbf{P}}_{l}}(\mathbf{y}-\mathbf{H}{{{\mathbf{\hat{u}}}}_{l}}), \\
&\upsilon _{l}^{2}=\frac{{\left\| \mathbf{\hat y}-\mathbf{H}{{\mathbf{\hat{u}}}_{l}} \right\|}^2-M\sigma _{\omega '}^{2}}{\rm tr({{\mathbf{H}}^{H}}\mathbf{H})}, \\
&{{\tau }^{2}}=\frac{1}{2N}\mathrm {tr}({{\mathbf{D}}_{l}}\mathbf{D}_{l}^{H}){\upsilon} _{l}^{2}+\frac{\gamma _{l}^{2}}{4N}\mathrm {tr}({{\mathbf{P}}_{l}}\mathbf{P}_{l}^{H})\sigma _{\omega ' }^{2},\\
&{{{\mathbf{{\hat{u}}}}}_{l+1}}=\mathbb{E}\{\mathbf{u}|{{\mathbf{r}}_{l}},{{\tau }_{l}}\},
\end{align}
where ${{\mathbf{D}}_{l}}=\mathbf{I}-\gamma _{l}^{{}}{{\mathbf{P}}_{l}}\mathbf{H}$. {Two trainable variables in each layer, $ (\lambda _{l}^{{}},\gamma _{l}^{{}})$, are introduced in the OAMP-NET, which is the difference from the TISTA in \cite{TISTA}. If ${\lambda _{l}^{{}}=\gamma _{l}^{{}}}$, the OAMP-NET is simplified to the TISTA. } The transmitted symbol, $\mathbf{u}$, is obtained from the real alphabet modulation set, $\mathcal{\tilde{A}}=\{{{a}_{1}},...,{{a}_{m}},...,{{a}_{\sqrt{M}}}\}$ and the corresponding posterior mean estimator, $\mathbb{E}\{\mathbf{u}|{{\mathbf{r}}_{l}},{{\tau }_{l}}\}$, in (17) for each element of ${{{\mathbf{{\hat{u}}}}}_{l+1}}$ is given by
%where ${{\mathbf{D}}_{l}}=\mathbf{I}-\gamma _{l}^{{}}{{\mathbf{P}}_{l}}\mathbf{H}$. The transmitted symbol, $\mathbf{u}$, is obtained from the alphabet modulation set, $\mathcal{A}=\{{{a}_{1}},...,{{a}_{m}},...,{{a}_{{M}}}\}$ and the corresponding posterior mean estimator, $\mathbb{E}\{\mathbf{u}|{{\mathbf{r}}_{l}},{{\tau }_{l}}\}$, in (17) for each element of ${{{\mathbf{{\hat{u}}}}}_{l+1}}$ is given by
\begin{equation}\label{3}
  \mathbb{E}\{{{{{{u}^{n}}}}}|{r}_{l}^{n},{{\tau }_{l}}\}\text{=}\frac{\sum\nolimits_{{{a}_{m}\in \mathcal{A}}}{{{a}_{m}}}\mathcal{N}({{a}_{m}};{r}_{l}^{n},\tau _{l}^{2})}{\sum\nolimits_{{{a}_{m}\in \mathcal{A}}}{\mathcal{N}({{a}_{m}};{r}_{l}^{n},\tau _{l}^{2})}}.
\end{equation}
Different from the OAMP algorithm, the damping parameter $\beta$ is removed from OAMP-NET.

Each layer in Fig. \ref{figOAMPNet} contains only two adjustable variables $ (\lambda _{l}^{{}},\gamma _{l}^{{}})$. Therefore, the total number of trainable variables is equal to $2L$ if there are $L$ layers.
Furthermore, the number of trainable variables of the OAMP-NET is independent of the number of subcarriers $N$ in OFDM and is only related to by the number of layers $L$. This feature is advantageous for an OFDM system with many subcarriers. The trainable variables of the OAMP-NET are much fewer than those of FC-DNN \cite{8052521liye} and ComNet \cite{gao2018comnet}. Moreover, convergence, stability, and speed can be improved during training.

\subsection{Complexity Analysis}
%The complexity of model-based aided receivers and AI-aided receivers are concluded in Table \ref{tableComplexity} in terms of the amount of floating-point multiplication-adds (FLOPs), memory usage, and running time required to complete a single-forward pass of one OFDM block. \textbf{LS+OFDM} represents traditional LS channel estimation and OFDM detection which is the result of received data divided by channel in frequency domain. The traditional LMMSE channel estimation and OFDM detection is denoted by \textbf{LMMSE+OFDM}. All listed algorithms above CE-NET+OAMP in Table \ref{tableComplexity} is model-based, others are AI-aided. \textbf{CE-NET+OAMP} denotes that the CSI is obtained by CE-NET and the signal is detected by the OAMP algorithm. Our proposed AI-aided receiver is noted as AI receiver and other AI-aided receivers are denoted as FC-DNN \cite{8052521liye} and ComNet \cite{gao2018comnet}.

The complexity of the model-based receiver and the AI-aided receiver is compared in Table \ref{tableComplexity} in terms of the number of floating-point multiplication-adds (FLOPs), the number of parameters, and running time required to complete a single-forward pass of one OFDM block. {\textbf{LS+OFDM} represents the traditional LS channel estimation and expert OFDM detection which is the result of $\mathbf Y_{\rm {D}}$ divided by estimated channel $\hat{\mathbf H}_{\text {LS}}$}. $\mathbf{Y}_{{\rm {D}}}$ represents the received signal in frequency domain. Traditional LMMSE channel estimation and expert OFDM detection is denoted by \textbf{LMMSE+OFDM}. All algorithms listed before CE-NET+OAMP in Table \ref{tableComplexity} are model-based, whereas others are AI-aided. \textbf{CE-NET+OAMP} indicates that the CSI is obtained by CE-NET and the signal is detected by the OAMP algorithm. Our proposed AI-aided receiver is designated as AI receiver and other AI-aided receivers are denoted as \cite{8052521liye} and ComNet \cite{gao2018comnet}.

%For the OAMP algorithm and OAMP-NET, the complexity is dominated by the computation of the covariance matrix in (\ref{w_hat}) whose complexity is $\mathcal{O}({{n}^{3}}L)$. The number of FLOPs of OAMP algorithm is about $1.05\times 10^{7}$ ($L=5$). From the Fig. \ref{figOAMPNet}, the total number of trainable variables is equal to $2L$, since each layer of the OAMP-Net contains only two adjustable variables $ (\lambda _{l}^{{}},\gamma _{l}^{{}})$. Furthermore, the number of trainable variables of the OAMP-Net is independent of the number of subcarriers $N$, and only determined by the number of layers $L$. This is an advantageous feature for largescale OFDM system.

For the OAMP algorithm and the OAMP-NET, complexity is dominated by the computation of the covariance matrix in (\ref{w_hat}) whose complexity is $\mathcal{O}({{n}^{3}}L)$. The number of FLOPs of OAMP algorithm is approximately $1.05\times 10^{7}$ ($L=5$). From Fig. \ref{figOAMPNet}, the total number of trainable variables is equal to $2L$, as each layer of the OAMP-Net contains only two adjustable variables $ (\lambda _{l}^{{}},\gamma _{l}^{{}})$. %Furthermore, the number of trainable variables of the OAMP-Net is independent of the number of subcarriers $N$, and is only determined by the number of layers $L$. This feature is advantageous for large-scale OFDM systems.

From Table \ref{tableComplexity}, the LMMSE+OAMP requires the largest number of FLOPs and the LS+OFDM requires the least. For the model-based receivers, the number of FLOPs and running time tend to increase from topmost item downwards, in which the LS+OFDM needs ${3.3\times10^{-3}}$ s, whereas LMMSE+OAMP needs ${5.6\times10^{-2}}$ s . There is a gap between the FLOPs of LMMSE and CE-NET because LMMSE needs to compute the inverse matrix as in (4) while CE-NET has been trained. Therefore, the difference of the FLOPs between LMMSE+OAMP and CE-NET+OAMP is approximately 1.6 million FLOPs. Obviously, our proposed AI receiver demands the same number of FLOPs as LS+OAMP as the AI receiver does not change the structure of the OAMP algorithm and merely introduces ${6.5\times10^{-2}}$ Mbytes in training parameters. For the DNN, the number of parameters is used to measure the complexity. Note that the FC-DNN has the most parameters followed by ComNet and then by our AI receiver has the least. On the contrary, the running time of the AI receiver is much more than that of FC-DNN and ComNet since the computation of the inverse matrix in the OAMP-NET structure consumes lots of time. The running time of the model-driven and model-based receivers are generally a little longer than those of completely data-driven AI-aided receivers. The time consumption is influenced not only by FLOPs but also the limitations of intensity and roofline [9], the platform, cache size, and realization method. The presented running time of the model-based methods and the AI receiver are computed by MATLAB, whereas the FC-DNN and ComNet are tested on Python with parallelization.
\begin{table}[!h]
		\centering
		\footnotesize
        \caption{Complexity comparison for proposed schemes and competing methods}	
		\begin{tabular}{>{\sf }lllll}    %lcrrr
			\toprule
			Algorithm & FLOPs &  \#parameter & Time  \\
			\midrule
			LS+OFDM &768  & $\times$   &  ${3.3\times 10^{-3}}$ s\\	
			%\hline
			LMMSE+OFDM	&1.6M   & $\times$   & $1.3\times 10^{-2}$ s \\
			%\hline
			LS+OAMP	&10.5M   &$\times$  & $5.0\times 10^{-2}$ s \\
            {LMMSE+OAMP}	&12.1M   &$\times$   & $5.6\times 10^{-2}$ s \\
            {CE-NET+OAMP}	&10.5M   &0.065MBytes  & $5.1\times 10^{-2}$ s \\
            {AI receiver}	&10.5M   &0.065MBytes  & $5.1\times 10^{-2}$ s \\
            {FC-DNN}	&$\times$   & 9.30MBytes & $1.2\times 10^{-6}$ s  \\
            {ComNet}	&$\times$  & 2.4MBytes   & $7.2\times 10^{-6}$ s\\
			\bottomrule
		\end{tabular}
		\label{tableComplexity}
	\end{table}
\section{Simulation Results and Discussions}
%In this section, the simulated performance and corresponding discussions are presented. Firstly, the simulation parameters are illustrated, including channel conditions and the setting of parameters. Secondly, the performance of AI receiver is evaluated. Thirdly, the robustness of OAMP-NET is analyzed.

This section presents simulated results, which provide guideline for the OTA test of AI-aided receiver for a CP-free OFDM system.
\subsection{Parameters Setting}
\begin{figure}[!h]
\centering
\includegraphics[width=3.5in]{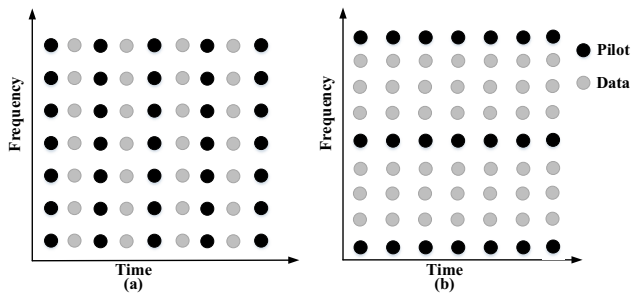}
\caption{Pilot arrangement of the quasi-static fading channels and the fast fading channels. (a) Continuous pilot arrangement for the quasi-static fading channels. (b) Comb type pilot arrangement of the fast fading channels.}
\label{figpilot}
\end{figure}

A CP-free OFDM system is with 64 subcarriers in our simulation. To estimate the channel, different types of pilots are inserted into the OFDM blocks according to the speed of channel variation. For the quasi-static fading channels that changes after passing several OFDM blocks, the pilot OFDM block and subsequent several data OFDM blocks usually form a frame. In our simulation, one pilot OFDM block and one data OFDM block constitute one frame as shown in Fig. \ref{figpilot}(a) for simplicity. For the fasting fading channels that changes between adjacent OFDM blocks, comb type pilots are uniformly distributed within each OFDM block as in Fig. \ref{figpilot}(b).

 To verify the robustness of the proposed algorithms, two types of channel models are simulated: the wireless world initiative new radio II (WINNNERII) \cite{2007WINNERII} and Stanford University Interim (SUI) \cite{cho2010mimo}. In the WINNNERII channel, the carrier frequency is 2.6 GHz, the number of paths is 24, and typical urban channels with max delay 16 samples are used, which is also consistent with the channel models in \cite{8052521liye} and \cite{gao2018comnet}. For the SUI channel, the delay spreads are [0, 4, 10] and the corresponding average powers of the three paths are [0 dB, -5 dB, -10 dB]. The types of modulation are 16-QAM and 64-QAM.

The CE-NET and OAMP-NET are trained by minimizing the cost between predictions and actual labels by using the adaptive moment estimator (Adam) optimizer. The learning rate is set to 0.001. We select the $\ell_2$ loss as the cost function. In the CE-NET, the training and testing sets contain ${3, 000, 000}$ and ${1, 000, 000}$ channel samples, respectively. The batch size and epochs are set to 50 and 2,000, respectively.
The CE-NET and the OAMP-NET are trained separately. After finishing the training of the CE-NET, the OAMP-NET is trained by known data symbols with $10, 000$ epochs. The OAMP-NET has 10 layers. At each epoch, the training and development sets both contain 1,000 samples, respectively. We continue generating the test data for the OAMP-NET until the number of bit errors exceeds 1,000. The training parameters of the CE-NET and the OAMP-NET are summarized in Tables \ref{train} and \ref{train1}, respectively.
%$\left[ 0\text{ }dB,-5dB,-10dB \right]$      $\left[ 0,4,10\right]$
%In SUI channel, the delay spread chooses $\left[ 0,4,10\right]$ and power profile uses corresponding $\left[ 0\text{ }dB,-5dB,-10dB \right]$. The types of modulations are 16-QAM and 64-QAM. CE-NET and OAMP-NET are trained by minimizing the cost between predictions and actual labels by using the adaptive moment estimator (Adam) optimizer. The learning rate is set to 0.001. For cost function, we select the $\ell_2$ loss. In CE-NET, the training and testing sets contain 3,000,000 and 1,000,000 samples, respectively. The batch size and epochs are set to 50 and 2,000, respectively.

%The CE-NET is initially trained with a time-varying channel and OAMP-NET is then trained with 10,000 epochs. The OAMP-NET has 10 layers. The batch size of OAMP-NET is set to 1,000. At each epoch, the training and development sets contain 1,000 and 1,000 samples, respectively. We keep on generating the test data for the OAMP-NET until the number of bit errors exceeds 1,000. The training parameters of CE-NET and OAMP-NET are summarized in Table \ref{train} and Table \ref{train1}, respectively.

	\begin{table}[!h]
		\centering
		\caption{Training Parameters of CE-NET. }	
%The parameters in the AI aided OFDM receivers need to be trained through labeled data in advance. This table presents the choice of training parameters in simulations.
		\footnotesize
		\begin{tabular}{>{\sf }l|l}    %
			\toprule
			Parameter&   Value   \\
			\midrule
			SNR&   5-40 dB   \\
			Loss function&	  $\ell_2$ \\
            Batch size&	  50 \\
			Epoch&	 2000 \\
			Initial learning rate&	  0.001 \\
			Optimizer&	  Adam \\
			\bottomrule
		\end{tabular}
		\label{train}
	\end{table}
	\begin{table}[!h]
		\centering
		\caption{Training Parameters of OAMP-NET. }	
%The parameters in the AI aided OFDM receivers need to be trained through labeled data in advance. This table presents the choice of training parameters in simulations.
		\footnotesize
		\begin{tabular}{>{\sf }l|l}    %
			\toprule
			Parameter&   Value   \\
			\midrule
			SNR&   5-40 dB   \\
			Loss function&	 $\ell_2$ \\
            Batch size&	  1000 \\
			Epoch&	 10000 \\
			Initial learning rate&	  0.001 \\
			Optimizer&	  Adam \\
			\bottomrule
		\end{tabular}
		\label{train1}
	\end{table}

The following naming conventions are employed to concisely present the performance:
\begin{itemize}
\item (XX dB): obtained by training a neural network with the received data under SNR = XX dB, but predicted the results among different SNRs;
%\item (40dB): the same as  (20dB) except that training SNR equals 40dB.
\item (ALL dB): indicates that the neural network is trained under the matched SNR value. For example, the predicted value under SNR = 25 dB is obtained when the neural network is also trained under SNR = 25 dB;
\item OAMP-NET(xxQAM, XXdB): the specific OAMP-NET is trained under xx-QAM with XX dB and trained network structure is used to predict the results of other situations;
\item 16 pilot: comb pilots with 16 pilots applied;
\item 64 pilot: continuous pilots with 64 pilots applied;
\item S: the SUI channel;
\item W: the WINNERII channel; and
\item S/W: the SUI or WINNERII channel.
\end{itemize}

%•	(XXdB): obtained by training a neural network with the received data under SNR = XXdB, but predicted the results among the SNRs of 5, 10, 15, 20, 25, 30, 35 and 40 dB;
%•	(ALLdB): indicates that the neural network is trained under the matched SNR value. For example, the predicted value under SNR = 25 dB is obtained when the neural network is also trained under SNR = 25 dB;
%•	OAMP-NET(xxQAM, XXdB): the specific OAMP-NET is trained under xx-QAM with XXdB and trained network structure is used to predict the results of other situations;
%•	16pilot: comb pilots with 16 pilots applied;
%•	64pilot: continuous pilots with 64 pilots are applied;
%•	S: the SUI channel;
%•	W: the WINNERII channel; and
%•	S/W: the SUI or WINNERII channel;

%For example, OAMP-NET(64QAM, ALLdB) represents that the specific  OAMP-NET is trained under 64-QAM with a SNR and the trained network structure is used to predict the results of other situations with matched SNR. If there are no special illustrations, the continuous pilots, ALLdB and  WINNERII channel are applied in the simulation.
For example, OAMP-NET(64QAM, ALLdB) indicates that the specific OAMP-NET is trained under 64-QAM with a SNR and the trained network structure is used to predict the results of other situations with matched SNR. If no special illustrations are present, then the continuous pilots, the ALLdB and WINNERII channels, are applied in the simulation.

\subsection{Performance Analysis of CE-NET}
\begin{figure}[!ht]
\centering
\subfigure[Continuous pilots on the WINNERII channel]{
\begin{minipage}[t]{0.45\textwidth}
\centering
\includegraphics[width=3.2in]{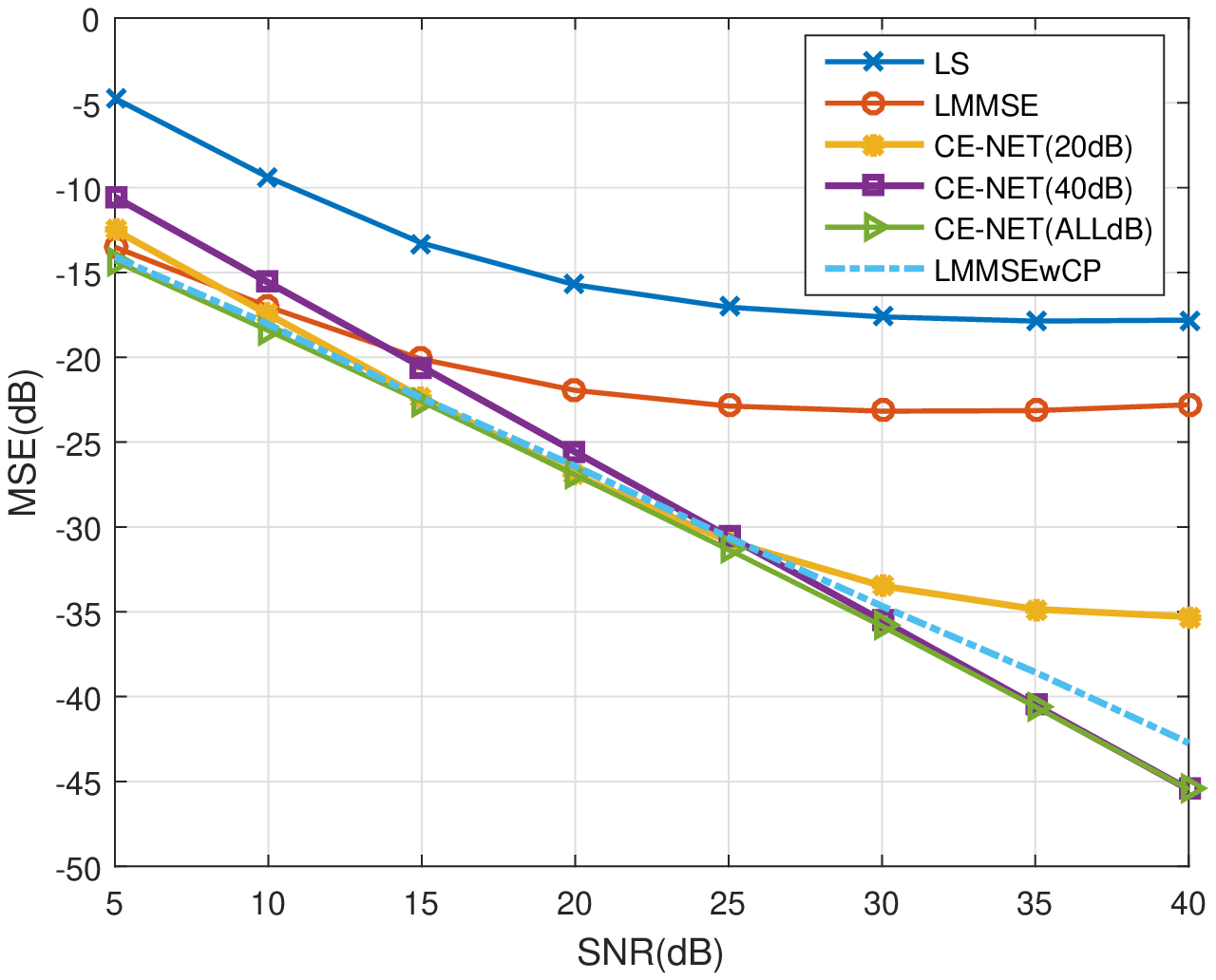}
%\caption{The impact of number of MTs.}
\label{figa}
\end{minipage}}
\subfigure[16 comb pilots on the WINNERII channel]{
\begin{minipage}[t]{0.45\textwidth}
\centering
\includegraphics[width=3.2in]{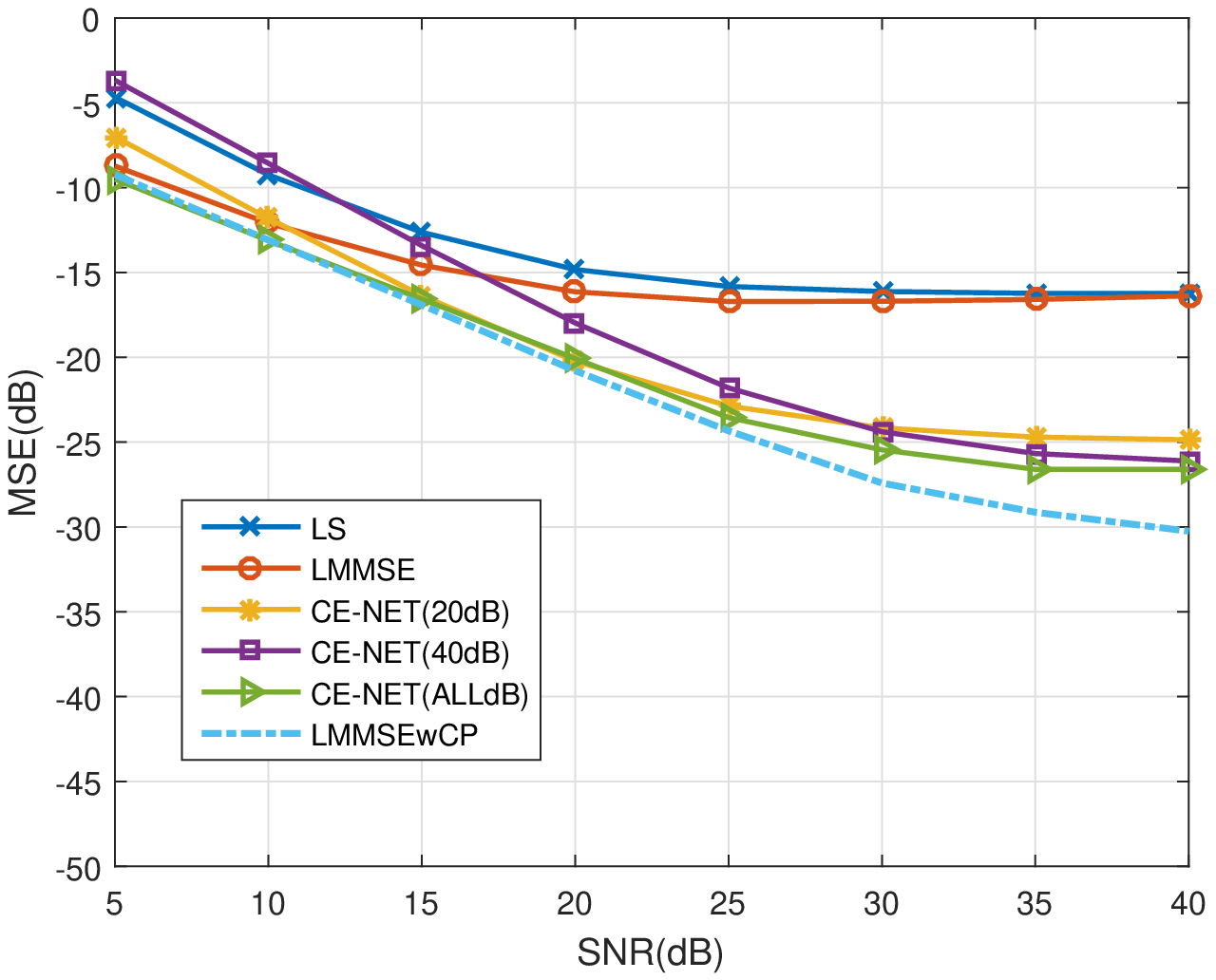}
%\caption{The impact of maximum tolerable delay.}
\label{figb}
\end{minipage}}

\caption{MSE curves of CE-NET and traditional methods in different conditions. (a) continuous pilots on WINNERII channel. (b) comb pilots on WINNERII channel.}
\end{figure}
The purpose of the CE-NET is to acquire accurate CSI information, which is beneficial for the subsequent signal detection utilizing OAMP-NET and adopting high-order QAM modulation. With only a single training, the CE-NET can adapt to time-varying channels.

%Fig. \ref{figa} shows the mean square error (MSE) performance of different channel estimation methods applying to WINNERII channel.  As shown in Fig. \ref{figa}, the MSE curves of CE-NET(20dB) present saturation property like traditional channel estimation methods of LS and LMMSE. However, the CE-NET(40dB) rectifies the nonlinear effects and have better performance than LMMSE with sufficient CP when SNR is higher than 25dB, which suggests its effectiveness and feasibility. Furthermore, the MSE of CE-NET(ALLdB) is lower than LMMSE with sufficient CP. Compared with the CE-NET(ALLdB), CE-NET(40dB) has a little performance loss but endures less training complexity.

Fig. \ref{figa} shows {the MSE} performance of different channel estimation methods applied to the WINNERII channel. As shown in Fig. \ref{figa}, the MSE curves of CE-NET(20 dB) present a saturation property similar to the traditional channel estimation methods of LS and LMMSE. However, the CE-NET(40 dB) rectifies the nonlinear effects and has better performance than LMMSE with sufficient CP when SNR is higher than 25 dB, which suggests its effectiveness and feasibility. Furthermore, the MSE of CE-NET(ALLdB) is lower than LMMSE with sufficient CP. Compared with the CE-NET(ALLdB), CE-NET(40 dB) has minimal performance loss but endures less training complexity.

%As shown in Fig. \ref{figb}, the CE-NET(20dB) is more robust than that of CE-NET(40dB) which works effectively in low SNR. In addition, CE-NET(ALLdB) consumes more training cost but achieves only little gain. Therefore, in this scenario, we choose the weights and bias for CE-NET trained under the 20dB. We can also observe that the performance of CE-NET working in comb pilots is much better than traditional LS and LMMSE and obtain approximately 10 dB gain in the SNR of 30 dB. In addition, the biggest gap between CE-NET(20dB) and LMMSE with sufficient CP called LMMSEwCP in the Fig. \ref{figb} is about 7 dB in the SNR of 40 dB whereas the least difference is lower than 1 dB when the SNR is 20 dB. Overall, the CE-NET is effective and robust since the performance is greatly beyond LS and LMMSE.

As shown in Fig. \ref{figb}, the CE-NET(20 dB) is more robust than that of CE-NET(40 dB) which works effectively in low SNR. In addition, CE-NET(ALLdB) entails more training cost but achieves only little gain. Therefore, in this scenario, we choose the weights and bias for the CE-NET trained under 20 dB. Note also that the performance of the CE-NET working in comb pilots is much better than the traditional LS and LMMSE and obtain approximately 10 dB gain when SNR = 30 dB. In addition, the biggest gap between CE-NET(20 dB) and LMMSE with sufficient CP called LMMSEwCP in the Fig. \ref{figb} is approximately 7 dB at SNR = 40 dB, but the least difference is lower than 1 dB at SNR = 20 dB. Overall, the CE-NET is effective and robust as it significantly outperforms LS and LMMSE.

For the SUI channels, the channel estimation performance is similar with that of the WINNERII channels, which also suggests the effectiveness of the CE-NET. Compared with the WINNERII channels, the difference of the SUI channels lies in lower delay and fewer paths. By applying the AI receiver to the new channel, the adaptation and flexibility for different channel environments have been demonstrated from the above discussion.

\subsection{Performance of OAMP-NET}
\subsubsection{Impact of the number of Iterations}
%The complexity and the number of training parameters of AI receiver largely depends on the number of layers of OAMP-NET. In order to explore the appropriate number of layers, different simulations are executed by setting different parameters. Table \ref{tableBer} and Table \ref{tableSer} shows the BER and symbol error rate (SER) comparison between OAMP algorithm and OAMP-NET in different SNR with 16-QAM, respectively. The CSIs of two detection methods are both estimated by CE-NET. When $L=1$, both OAMP and OAMP-NET have poor performance whether the SNR is 25 dB or 40 dB. With the increasing of iterations, the BER and SER become lower whether the detection method is OAMP or OAMP-NET. When $L=3$, we can see that the BER decrease quickly compared with $L=1$, from 6.516e-3 to 7.34e-4 in 40dB, approximately reducing 10 times. When the iteration times change from 3 to 5, the performance increase can achieve from 5.46e-4 to 9.07e-4. However, when the iteration times change from 5 to 10, the growth of BER performance becomes limited just increasing approximately 4.0e-5 in 40dB, which are is at the expense of doubled computation complexity. It has the tendency to decrease slowly on the BER when the iterations become large. In addition, in the same iterations, the BERs in 40 dB are always lower than those in 25 dB.  We can also observe from Table \ref{tableBer} that the OAMP-NET can improve the BER performance of OAMP algorithm greatly, such as from 1.57e-3 to 1.88e-4 in 40 dB for $L=5$ decreasing almost 10 times.

The complexity and the number of training parameters of the AI receiver largely depend on the number OAMP-NET layers. To explore the appropriate number of layers, different simulation settings are executed. Table \ref{tableBer} shows the BER comparison between the OAMP algorithm and the OAMP-NET in different SNR with 16-QAM, respectively. The CSI of the two detection methods is estimated by the CE-NET. When $L=1$, both the OAMP and the OAMP-NET have poor performance when the SNR is 25 dB or 40 dB. With increasing iterations, the BER reduces for both methods. When $L=3$, the BER decreases quickly compared to when $L=1$, from ${6.5\times10^{-3}}$ to ${7.3\times10^{-4}}$ at SNR = 40 dB. When the number of the iterations increases from 3 to 5, the BER decreased to ${1.9\times10^{-4}}$ at SNR = 40 dB. {However, when the iteration number changes from 5 to 10, the decrease of BER becomes limited, from  ${2.9\times10^{-3}}$ to ${2.6\times10^{-3}}$ at SNR = 20 dB, but the computation complexity is doubled.}  The BER has the tendency to decrease slowly when the iterations become large. In addition, in the same iterations, the BERs in 40 dB are always lower than those in 25 dB. Table \ref{tableBer} indicates that the OAMP-NET can considerably improve the BER performance of the OAMP algorithm, such as from $1.6\times10^{-3}$ to {$1.9\times10^{-4}$} in 40 dB for $L=5$, decreasing by almost 10 times.
%{$1.88\times10^{-4}$} $L=5$
\begin{table}[!h]
		\centering
		\footnotesize
        \caption{BER comparison between OAMP and OAMP-NET with 16-QAM}	
		\begin{tabular}{>{\sf }lllll}    %lcrrr
			\toprule
			Iteration & OAMP&  OAMP-NET  & OAMP & OAMP-NET \\
                      &(25dB)&  (25dB)  & (40dB) & (40dB) \\
			\midrule
			$L$=1 &{$1.1\times10^{-2}$}	&{$1.1\times10^{-2}$}	&{$7.9\times10^{-3}$}	&{$6.5\times10^{-3}$}  \\	
			%\hline
			$L$=3	&{$6.2\times10^{-3}$}	&{$3.8\times10^{-3}$}	&{$2.2\times10^{-3}$}	&{$7.3\times10^{-4}$}  \\
			%\hline
			$L$=5	&{$5.4\times10^{-3}$}	&{$2.9\times10^{-3}$}	&{$1.6\times10^{-3}$}	&{$1.9\times10^{-4}$} \\
            $L$=10 &{$5.1\times10^{-3}$}	&{$2.6\times10^{-3}$}	&{$1.1\times10^{-3}$}	&{$1.5\times10^{-4}$} \\
			\bottomrule
		\end{tabular}
		\label{tableBer}
	\end{table}

\subsubsection{Performance for the WINNERII and SUI channels with perfect CSI}
\begin{figure}[!ht]
\centering
\subfigure[64-QAM on WINNERII channel.]{
\begin{minipage}[b]{0.45\textwidth}
\centering
\includegraphics[width=3.2in]{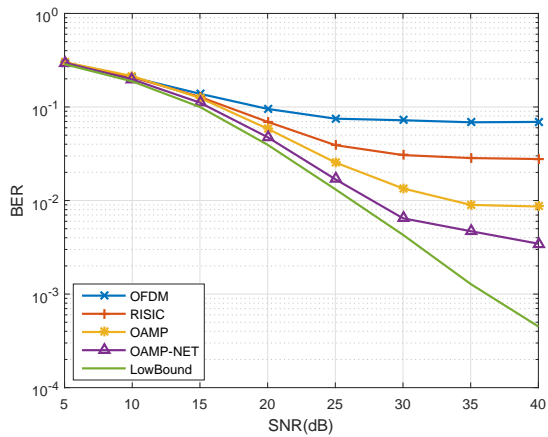}
\label{figOAMP64WINNERII}
\end{minipage}
}
\subfigure[16-QAM on WINNERII channel.]{
\begin{minipage}[b]{0.45\textwidth}
\centering
\includegraphics[width=3.2in]{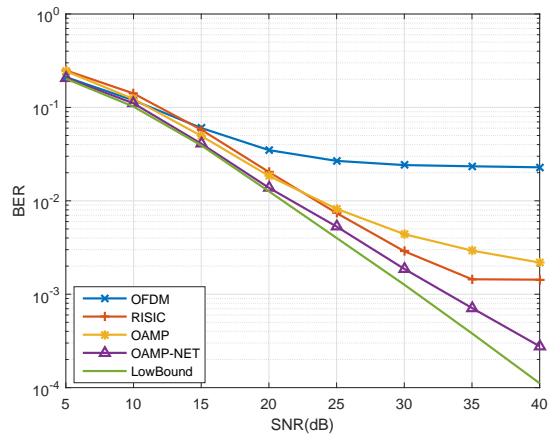}
\label{figOAMP16WINNERII}
\end{minipage}
}
\label{figWINNERII}
 \caption{BER curve of OAMP-NET and competitive methods under the CP-free case with 64-QAM and 16-QAM on WINNERII channel. (a) 64-QAM. (b) 16-QAM.} \label{winner2}
\end{figure}

The BER curves of the OAMP-NET and the competitive methods in the CP-free case with 64-QAM and 16-QAM are shown in Fig. \ref {winner2}. In the simulation below, the perfect CSI is employed for signal detection at the receiver. \textbf{LowBound} indicates that the maximum likelihood (ML) detector is used in the conventional CP-OFDM system.

%In Fig. \ref{figOAMP64WINNERII}, classical OFDM method do not satisfy orthogonal property of subcarriers due to the effect of removing CP. Thus, the method exhibits poor performance. On one hand, the BER of OAMP-NET is lower than that of the RISIC algorithm in \cite{1998Kim}. On the other hand, the OAMP-NET can remarkably improve the performance of the OAMP algorithm. These findings indicate the superiority of OAMP-NET only by introducing a few parameters. In addition, the gap between LowBound and the OAMP-NET is extremely small when the SNR is lower than 30 dB. However, the gap becomes slightly large when the SNR is higher than 30 dB.

In Fig. \ref{figOAMP64WINNERII}, the traditional OFDM detection does not satisfy the orthogonal property of subcarriers because of the effect of CP removal. Thus, the method exhibits poor performance. On the one hand, the BER of the OAMP-NET is lower than that of the RISIC algorithm in \cite{1998Kim}. On the other hand, the OAMP-NET can remarkably improve the performance of the OAMP algorithm. These findings indicate the superiority of the OAMP-NET by introducing only a few parameters. In addition, the gap between LowBound and the OAMP-NET is extremely small when the SNR is lower than 30 dB. However, the gap becomes slightly larger when the SNR is higher than 30 dB.

%Figure \ref{figOAMP16WINNERII} shows the BER performance for 16-QAM modulation in WINNERII channel. Different from that with 64-QAM, the BER of the OAMP algorithm for 16-QAM is higher than that of the RISIC algorithm when the SNR is over 25 dB. However, the BER of OAMP-NET can be considerably decreased and much lower than that of the RISIC algorithm. The performance improvement is because the parameters $(\lambda _{l}^{{}},\gamma _{l}^{{}})$ can be tuned in the OAMP-NET in each layer, resulting in a flexible network. In addition, the gap between OAMP-NET and LowBound is smaller than that with 64-QAM. When the SNR is blow 20 dB, the BER of OAMP-NET can approach the performance of LowBound. With the mounting SNR, the gap becomes a little larger but far less than that with 64-QAM. In summary, the OAMP-NET exhibits excellent BER performance compared with the existing algorithms and is close to the performance limit of an OFDM system with sufficient CP for 16-QAM and 64-QAM. However, the performance of OAMP-NET with 16-QAM is more accessible to the performance limit than that with 64-QAM which still has some improvement space.
Figure \ref{figOAMP16WINNERII} shows the BER performance for 16-QAM modulation in the WINNERII channel. Unlike that with 64-QAM, the BER of the OAMP algorithm for 16-QAM is higher than that of the RISIC algorithm when the SNR is over 25 dB. However, the BER of the OAMP-NET is much lower than that of the RISIC algorithm since the parameters $(\lambda _{l}^{{}},\gamma _{l}^{{}})$ are tuned in the OAMP-NET in each layer, thereby resulting in a flexible network. In addition, the gap between the OAMP-NET and LowBound is smaller than that with 64-QAM. When the SNR is below 20 dB, the BER of the OAMP-NET approaches the performance of LowBound. With the mounting SNR, the gap becomes a little larger but much less than that with 64-QAM. In summary, the OAMP-NET exhibits excellent BER performance compared with the existing algorithms and is close to the performance limit of an OFDM system with sufficient CP for 16-QAM and 64-QAM. However, the performance of the OAMP-NET with 16-QAM is closer to the performance limit than that with 64-QAM. Therefore, there is still a room to improve a CP-free OFDM with 64-QAM modulation.

The BER performance of the relevant algorithms under the SUI channel is similar to those under the WINNERII counterpart. Different from the WINNERII channel, the BER of the OAMP algorithm for 16-QAM is always lower than that of the RISIC algorithm.

%The BER performance of relevant algorithms under the SUI channel is similar to those under the WINNERII counterpart, as shown in Fig. \ref{figOAMPSUI}. Figs. \ref{figOAMPSUI} is for 64-QAM and 16-QAM, respectively. Different from the WINNERII channel, the BER of the OAMP algorithm for 16-QAM is always lower than that of the RISIC algorithm.
%\begin{figure}[!h]
%\centering
%\subfigure[64-QAM on SUI channel.]{
%\begin{minipage}[b]{0.45\textwidth}
%\centering
%\includegraphics[width=3.2in]{SUI64QAM}
%\label{figOAMP64SUI}
%\end{minipage}
%}
%\subfigure[16-QAM on SUI channel.]{
%\begin{minipage}[b]{0.45\textwidth}
%\centering
%\includegraphics[width=3.2in]{SUI16QAM_1}
%\label{figOAMP16SUI}
%\end{minipage}
%}
%\label{figOAMPSUI}
% \caption{BER curve of OAMP-NET and competitive methods under the CP-free case with 64-QAM and 16-QAM on SUI channel. (a) 64-QAM. (b) 16-QAM.} \label{figOAMPSUI}
%\end{figure}

\subsection{Robustness Analysis}
\subsubsection{16 pilot}
\begin{figure}[!t]
\centering
\subfigure[16 comb pilots with 64-QAM.]{
\begin{minipage}[b]{0.45\textwidth}
\centering
\includegraphics[width=3.2in]{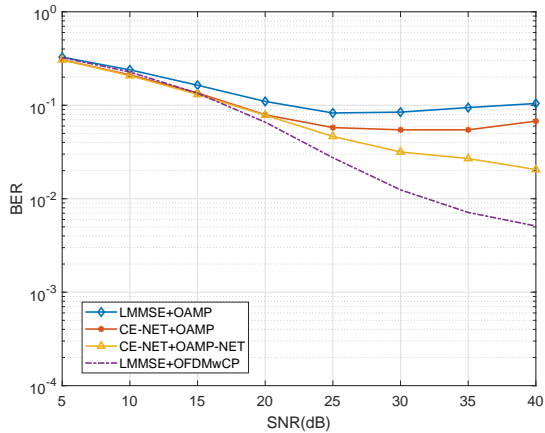}
\label{figcomb64}
\end{minipage}
}
\subfigure[16 comb pilots with 16-QAM.]{
\begin{minipage}[b]{0.45\textwidth}
\centering
\includegraphics[width=3.2in]{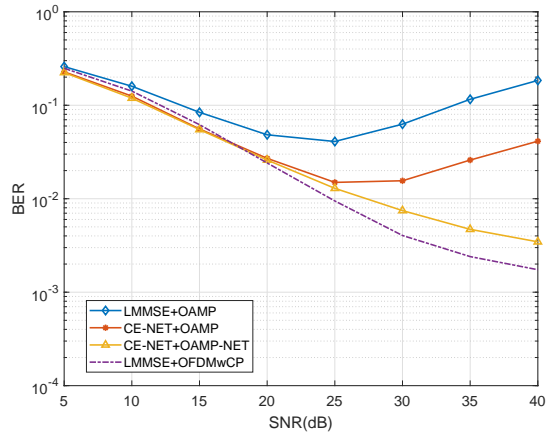}
\label{figcomb16}
\end{minipage}
}
%\label{figcomb}
 \caption{BER curve of OAMP and OAMP-NET with 16 comb pilots on the WINNERII channel. (a) 64-QAM. (b) 16-QAM.} \label{figcomb}
\end{figure}

%When the comb pilots are applied, the MSE of CE-NET has remarkable gap with continuous pilots as Fig. \ref{figb} shown. Fig. \ref{figcomb} shows the BER curves of OAMP and OAMP-NET in the case of 16 comb pilots on WINNERII channel.
When the comb pilots are applied, the MSE of CE-NET has a remarkable gap with the continuous pilots as Fig. \ref{figb} shows. Fig. \ref{figcomb} depicts the BER curves of the OAMP and the OAMP-NET in the case of 16 comb pilots on the WINNERII channel.

%We can see from Fig. \ref{figcomb64} that the model-based algorithm, LMMSE+OAMP, does not work since the LMMSE channel estimation in the comb pilots without CP is very poor. However, by introducing neural network, CE-NET+OAMP is better than LMMSE+OAMP where the BER decreases approximately by 0.02. However, both of them have the tendency to be saturated and the performance needs be improved further. In Fig. \ref{figcomb64}, the BER can be greatly decreased by applying OAMP-NET. The BER performance with 16-QAM is similar with 64-QAM. The difference from 64-QAM is that the LMMSE+OAMP and CE-NET+OAMP have more severe amplitude of fluctuation and bigger performance degradation when the SNR is more than 25dB, for which the reasons may be that the damping is exploded and the actual noise has mismatch with estimated noise since the first term of (7) introduces much more noise. It is apparent that the OAMP-NET has more great adaptation than the OAMP algorithm since the OAMP-NET can modify the noise mismatch and damping exploration, which also illustrates the advantages of OAMP-NET since it needs no damping parameter. Overall, Fig. \ref{figcomb} shows the performance of OAMP-NET in the case of imprecise channel estimation, and verifies the adaptation and robustness of it.

Fig. \ref{figcomb64} indicates that the model-based algorithm, LMMSE+OAMP, does not work as the LMMSE channel estimation in the comb pilots without CP. However, by introducing a neural network, CE-NET+OAMP outperforms LMMSE+OAMP even if both of them have the tendency to be saturated. In Fig. \ref{figcomb64}, the BER can be greatly decreased by applying the OAMP-NET. The BER performance with 16-QAM is similar to that with 64-QAM. The difference from 64-QAM is that the LMMSE+OAMP and CE-NET+OAMP have more severe amplitude fluctuation and bigger performance degradation when the SNR is more than 25 dB because the damping exploded and the actual noise does not match the estimated noise owing to the greater noise introduced by the first term of (7). {Clearly, the OAMP-NET is superior to the OAMP algorithm as the former needs no hand-crafted damping parameters but can prevent damping exploration.} Overall, the performance of the OAMP-NET in the case of imprecise channel estimation verifies its robustness.

\subsubsection{Adaptation of OAMP-NET}
%To test the robustness of OAMP-NET, as shown in Fig. \ref{figrobust}, the trained parameters of 64-QAM with 40dB in WINNERII is applied to other SNR of 64-QAM and 16-QAM. Fig. \ref {robust64} shows the BER performance of OAMP, OAMP-NET training under matched SNR and OAMP-NET training under 40dB in WINNERII channel. When SNR is lower than 15dB, the performance of OAMP-NET with 40dB is the same as that of matched SNR performance. With the increasing of the SNR, the performance experiences only a little loss. In addition, we can see that the OAMP-NET with 40dB perform much better than OAMP algorithm and only a little degradation compared with ALLdB, which suggests that OAMP-NET has great robustness and can adapt to different SNR with 64-QAM. To further verify the robustness, the trained parameters of 64-QAM with 40dB are applied to 16-QAM. Form Fig. \ref{robust16}, the OAMP-NET of 64-QAM with 40dB can work well on 16-QAM. On one hand, the OAMP-NET of 64-QAM with 40dB performs better than OAMP algorithm with 16-QAM when the SNR is higher than 15dB. The biggest BER gap can reach to 1.0e-3 when the SNR is 40 dB. On the other hand, the performance has little gap with the totally matched OAMP-NET, namely OAMP-NET of 16-QAM with ALLdB. In summary, the OAMP-NET which is trained in higher modulation mode can adapt to lower one, which is beneficial for modulation handover.
\begin{figure}[!t]
\centering
\subfigure[Robustness with 64-QAM.]{
\begin{minipage}[b]{0.45\textwidth}
\centering
\includegraphics[width=3.2in]{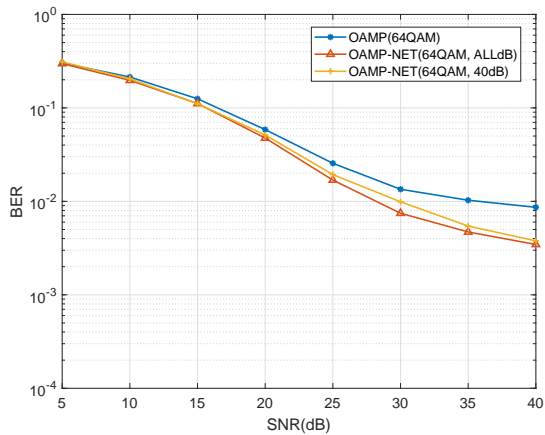}
\label{robust64}
\end{minipage}
}
\subfigure[Robustness with 16-QAM.]{
\begin{minipage}[b]{0.45\textwidth}
\centering
\includegraphics[width=3.2in]{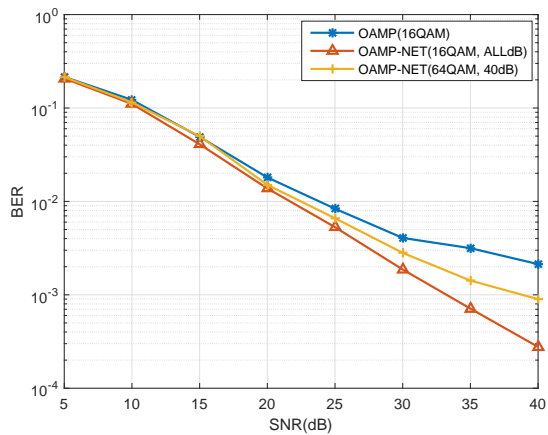}
\label{robust16}
\end{minipage}
}
%\label{figcomb}
 \caption{BER curve of OAMP-NET with different training cost on the WINNERII channel. (a) 64-QAM. (b) 16-QAM.} \label{figrobust}
\end{figure}
To test the robustness of the OAMP-NET, the trained parameters of 64-QAM with 40 dB in the WINNERII channels is applied to other SNRs of 64-QAM and 16-QAM, as shown in Fig. \ref{figrobust}. In Fig. \ref {robust64}, the BER performances of the OAMP, the OAMP-NET training under matched SNR, and OAMP-NET training under 40 dB in the WINNERII channels are depicted. When SNR is lower than 15 dB, the performance of the OAMP-NET with 40 dB is same as that of the matched SNR performance. With increasing SNR, the performance experiences only minimal loss. In addition, the OAMP-NET with 40 dB performs much better than the OAMP algorithm and shows only a little degradation compared with ALLdB, which suggests that OAMP-NET has strong robustness and can adapt to different SNRs with 64-QAM. To further verify the robustness, the trained parameters of 64-QAM with 40 dB are applied to 16-QAM in Fig. \ref{robust16}. Form the figure, the OAMP-NET of 64-QAM with 40 dB can work well on 16-QAM. On the one hand, the OAMP-NET of 64-QAM with 40 dB performs better than the OAMP algorithm with 16-QAM when the SNR is higher than 15 dB. On the other hand, the performance has little gap with the totally matched OAMP-NET, namely OAMP-NET of 16-QAM with ALLdB. In summary, the OAMP-NET trained in higher modulation mode can adapt to a lower one, which is beneficial for modulation handover.

%
%Table \ref{tablerobust} shows the BER of OAMP-NET trained with 40dB in different conditions including channel and modulation mode. OAMP-NET refers to the BER is obtained under corresponding matched conditions. OAMP-NET1 and OAMP-NET2 denote the specific OAMP-NETs trained under 16-QAM with 40dB in WINNERII and under 64-QAM with 40dB in WINNERII, respectively. From the results of OAMP-NET1, we can see that all BERs are lower than the OAMP algorithm. The phenomenon indicates that the OAMP-NET1 is not only valid to WINNERII channel, but also adapt to SUI channel since SUI channel is simpler than WINNERII channel. However, the BERs of OAMP-NET1 are a little higher than those of OAMP-NET, which is reasonable due to the total match for OAMP-NET. The similar results can be found for OAMP-NET2. For SUI channel with 64-QAM, OAMP-NET2 is more precise than OAMP-NET1. In contrast, for SUI channel with 16-QAM, OAMP-NET1 is more accurate than OAMP-NET2.

Table \ref{tablerobust} shows the BER of the OAMP-NET trained with 40 dB in different conditions, including the channel and modulation modes, where OAMP-NET refers to the BER obtained under corresponding matched conditions, OAMP-NET1 and OAMP-NET2 denote the specific OAMP-NETs trained under 16-QAM and under 64-QAM with 40 dB in the WINNERII channels, respectively. OAMP-NET1 results suggest that all BERs are lower than the OAMP algorithm. The phenomenon indicates that the OAMP-NET1 is not only valid for the WINNERII channels, but also adapt to the SUI channels as the latter is simpler than the former. However, the BERs of OAMP-NET1 are slightly higher than those of the OAMP-NET, which is reasonable due to the total match for the OAMP-NET. Similar results can be found for the OAMP-NET2. For the SUI channels with 64-QAM, OAMP-NET2 is more precise than OAMP-NET1. Conversely, for the SUI channels with 16-QAM, OAMP-NET1 is more accurate than OAMP-NET2.
\begin{table}[!h]
		\centering
		\footnotesize
        \caption{BER comparison between OAMP and OAMP-NET}	
		\begin{tabular}{>{\sf }lllll}    %lcrrr
			\toprule
			Iteration & 64-QAM    &  16-QAM       & 64-QAM & 16-QAM \\
                      &(W)&  (W)  & (S) & (S) \\
			\midrule
			OAMP        &0.010677	&0.002131	&0.010219	&0.002922  \\	
			%\hline
			OAMP-NET	&0.003458	&0.000277	&0.003654	&0.000148  \\
			%\hline
			%OAMP-NET1	&0.00931	&0.0022	    &0.0102	&0.0016 \\
%            OAMP-NET1 &0.0045       &0.000277	&0.0089	&0.00043 \\
%            OAMP-NET2 &0.003458	    &0.0008989	&0.0075	&0.0012 \\
            OAMP-NET1 &0.0045       &0.000277	&0.0089	&0.00043 \\
            OAMP-NET2 &0.003458	    &0.0008989	&0.0075	&0.0012 \\
			\bottomrule
		\end{tabular}
		\label{tablerobust}
	\end{table}
\subsection{Capacity Analysis}

%Fig. \ref{figcapacity} reveals that the capacity comparison between AI receiver and traditional OFDM with sufficient CP in different scenarios including continuous pilots and comb pilots, WINNERII and SUI channel. The bandwidth is 20 MHz in the simulation. We can see that the capacity of AI receiver with comb pilots in WINNERII or SUI channel performed best for all scenarios since the CP is removed from an OFDM block and comb pilots occupy less resource blocks of OFDM. Therefore, the spectrum efficiency is relatively effective and the amount of transmission data is significantly improved. It also accounts for that the curves of utilizing comb pilots are obviously beyond those of using continuous ones. Our proposed AI receiver can work well in these scenarios, which indicates that the AI receiver has significant efficiency on spectrum.

\begin{figure}[!t]
\centering
\includegraphics[width=3.2in]{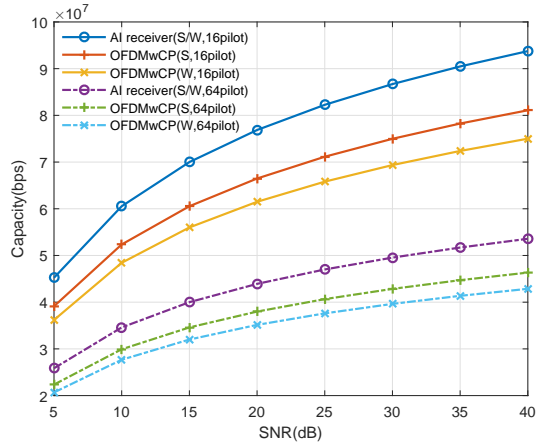}

\caption{Capacity comparison between AI receiver and traditional OFDM with sufficient CP in difference scenarios including the 64pilot and 16pilot, and the WINNERII and SUI channels.}
\label{figcapacity}
\end{figure}

Fig. \ref{figcapacity} compares capacities of the AI receiver and traditional OFDM with sufficient CP in different scenarios, including continuous pilots and comb pilots, as well as the WINNERII and SUI channels. The bandwidth in the simulation is 20 MHz. The capacity of the AI receiver with comb pilots in the WINNERII or SUI channels performed best for all scenarios because the CP is removed from an OFDM block and comb pilots occupy less resource blocks of OFDM. Therefore, the spectrum efficiency is relatively high and the amount of transmission data are significantly improved. It also accounts for the fact that the curves of the comb pilots are obviously beyond those of using continuous ones. Our proposed AI receiver can work well in these scenarios, which indicates that the AI receiver has significant spectrum efficiency.

%Figure \ref{figcomparison} compares the BER performance of AI receiver with other AI-aided methods, including FC-DNN studied by \cite{8052521liye} and ComNet of \cite{gao2018comnet}. FC-DNN become saturated when SNR is higher than 25dB, while the ComNet and AI receiver perform better in resolving the CP-free issue, which suggests the advantages of model-driven DNN. Compared with ComNet, the BERs of AI receiver is down greatly from 2.0e-2 to 6.0e-3 when SNR equals 30 dB, which suggests that OAMP-NET has the ability to recover raw bit stream more accurately.

\begin{figure}[!t]
\centering
\includegraphics[width=3.2in]{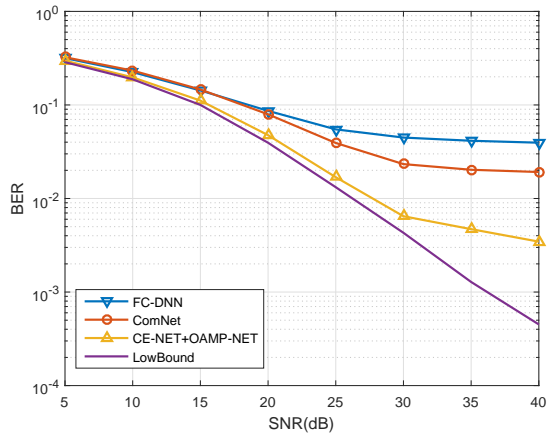}
% where an .eps filename suffix will be assumed under latex,
% and a .pdf suffix will be assumed for pdflatex; or what has been declared
% via \DeclareGraphicsExtensions.
\caption{BER performance of the AI receiver and other AI-aided methods with 64-QAM.}
\label{figcomparison}
\end{figure}

Fig. \ref{figcomparison} compares the BER performance of the AI receiver with other AI-aided methods, including the FC-DNN studied by \cite{8052521liye} and ComNet of \cite{gao2018comnet}. FC-DNN become saturated when the SNR is higher than 25 dB, whereas the ComNet and the AI receiver perform better in resolving the CP-free issue, which suggests the advantages of a model-driven DNN. Compared with those of the ComNet, the BER of the AI receiver reduces considerably from $2.0\times10^{-2}$ to $6.0\times10^{-3}$ when SNR equals 30 dB. Therefore, the OAMP-NET has the ability to recover raw bit stream more accurately.
\section{OTA Test and Result Discussion}
Apart from simulation, we have also developed a prototyping system to verify the effectiveness and feasibility of the proposed algorithms in real channel environments. In this subsection, we compare the performance of the AI receiver with the FC-DNN and ComNet receivers in OTA tests.

\subsection{System Setup}
\begin{figure}[!h]
		\centering
		\includegraphics[width = 3.5in]{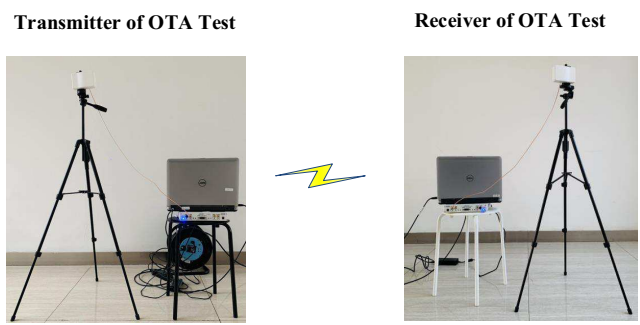}
		\caption{Real testing scenario consisting of transmitter and receiver. }
		\label{data}
	\end{figure}

As shown in Fig. \ref{data}, the real testing scenario includes a transmitter and a receiver of OTA, which offers processing and transmission of OFDM with 64 subcarriers. For simplicity, the modulation mode of QPSK is applied in the real scenario.  In \cite{yang2017rapro}, a novel 5G RaPro system is proposed to deploy FPGA-privileged modules on software defined radio (SDR) platforms. Such architecture has been proven to be flexible and scalable by deploying a multi-user full-dimension MIMO prototyping system.

%In this paper, we use the RaPro system as our testbed to test the OTA performance of AI receiver in CP-free OFDM like in \cite{2018jpw}. To implement CP-free OFDM communication system with the AI receiver, we employ two SDR nodes of the USRP RIO series manufactured by National Instruments. Each SDR node includes one RF transceiver of 20 MHz bandwidth and a programmable FPGA responsible for distributed signal processing, such as reciprocity calibration or OFDM (de)modulation \cite{yang2017rapro}.  Fig. \ref{ofdmsymbol} illustrates the OFDM block structure of the CP-free OFDM system. From Fig. \ref{ofdmsymbol}, each OFDM block contains 128 subcarries, where 64 subcarries are effective subcarriers for transmitting pilot symbols or data symbols, and 63 subcarries are used as guard band and a subcarrier as direct current (DC) offset. Each frame consists of one pilot OFDM block and one data OFDM block which is the same as simulation setting.

\begin{figure}[!ht]
		\centering
		\includegraphics[width = 2.5in]{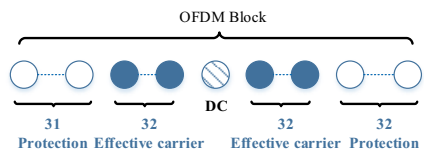}
		\caption{Structure of the OFDM blocks in a CP-free real testing scenario. }
		\label{ofdmsymbol}
	\end{figure}

We use the RaPro system as our testbed to evaluate the OTA performance of the AI receiver in a CP-free OFDM as found in \cite{2018jpw}. To implement a CP-free OFDM communication system with the AI receiver, we employ two SDR nodes of the universal software radio peripheral reconfigurable I/O (USRP-RIO) series manufactured by National Instruments. Each SDR node includes one RF transceiver of 20 MHz bandwidth and a programmable FPGA responsible for distributed signal processing, such as the reciprocity calibration or OFDM (de)modulation \cite{yang2017rapro}. Fig. \ref{ofdmsymbol} illustrates the OFDM block structure of the CP-free OFDM system. From Fig. \ref{ofdmsymbol}, each OFDM block contains 128 subcarriers, where 64 subcarriers are effective for transmitting pilot symbols or data symbols, 63 subcarriers are used as the guard band, and a subcarrier is the direct current (DC) offset. Each frame consists of one pilot OFDM block and one data OFDM block, which is the same as the simulation setting.

%For the transmitter of OTA test, there are 64 subcarries and 128 binary numbers as the QPSK is applied. We choose the first 16 binary numbers sequentially inserting $2^{16}$ data from $0000000000000000$ to $1111111111111111$, which can be regarded as labels to assess the BER for real channel environment. Then the pilot symbols are inserted and informed into a frame together with data symbols. Due to the existence of out-of-band emission, the bandwidth protection is needed what is to add guard band and DC offset according to the designed OFDM block. Next, the IFFT is performed and the frame head is added in order to detect the OFDM blocks precisely. Wireless signals are transmitted by an USRP-RIO through a RF antenna, whose center frequency is adjustable in the range of 1.2GHz-6GHz. The channel environment is indoor and the distance between the transmitted RF antenna and received RF antenna is 5m. The receiver of OTA test first conducts frame detection, then perform FFT and estimates the CSI, next transforms the CSI from frequency domain to time domain. Finally, the data signals are transmitted to our OAMP-NET to obtain the estimated transmitting bits, just as the Fig. \ref{figreceiver} shows.

For the transmitter in the OTA test, 64 subcarriers and 128 binary numbers as the QPSK are applied. We choose the first 16 binary numbers by sequentially inserting $2^{16}$ data from 0000000000000000 to 1111111111111111, which can be regarded as labels to assess the BER for the real channel environment. Then, the pilot symbols are inserted and packaged into a frame together with data symbols. Due to out-of-band emission, bandwidth protection requires the additional guard band and DC offset according to the designed OFDM block. Next, the IFFT is performed and the frame head is added to detect the OFDM blocks precisely. Wireless signals are transmitted by an USRP-RIO through a RF antenna, whose center frequency is adjustable in the range of 1.2{-}6 GHz. The channel environment is indoors and the distance between the transmitted RF antenna and received RF antenna is 5 meters. The receiver in the OTA test first conducts frame detection, then perform FFT and estimates the CSI, subsequently transforming the CSI from the frequency domain to the time domain. Finally, the data signals are transmitted to our OAMP-NET to obtain the estimated transmitting bits, just as Fig. \ref{figreceiver} shows.

%The proposed AI receiver development process can be divided into two phases, training phase and detecting phase. The training phase is developed in Python based on TensorFlow, relying on the GPUs' powerful computing ability. OTA data captured by USRP-RIO is used to train the weights and biases of the deep neural network via back propagation algorithm. These parameters are stored into a file after training and provided for the detecting phase.  In the detecting phase, the forward propagation is implemented in tensorflow, with the stored parameters as the initialization value of the weight matrices and bias vectors.

The development of the proposed AI receiver can be divided into two phases, the training and the detecting phases. The training phase is developed in Python based on TensorFlow by relying on the GPUs' powerful computing ability. OTA data captured by USRP-RIO is used to train the weights and biases of the DNN via the back propagation algorithm. These parameters are stored into a file after training and provided for the detecting phase. In the detecting phase, forward propagation is implemented in TensorFlow, with the stored parameters as the initialization value of the weight matrices and bias vectors.
\subsection{Experimental Results}

%We choose two different scenarios to test our AI receiver. Scenario 1 is the fixed indoor scenario, where the transmitter is five meters away from the receiver in the same room with fixed obstacles, windows, and walls around. Scenario 2 is the changing indoor scenario, in which the only difference with the Scenario 1 is that there exit some people walking around. Both two scenarios are relatively simple due to limited transmission distance, reflectors, and scatters, and that the corresponding real channels are similar to the exponential power delay profile defined in IEEE 802.11b \cite{cho2010mimo} which is denoted as exponential (EXP) channel. Therefore, our proposed AI receiver and FC-DNN as well as ComNet is trained offline under the EXP channel model to perform the OTA test, under the same SNR, by specific transmitter antenna gain.

We choose two different scenarios to test our AI receiver. Scenario 1 is the fixed indoor scenario, where the transmitter is 5 meters away from the receiver in the same room with fixed obstacles, windows, and walls. Scenario 2 is the changing indoor scenario, in which the only difference with Scenario 1 is the presence of people walking around. Both two scenarios are relatively simple because of limited transmission distance, reflectors, and scatters. Moreover, the corresponding real channels are similar to the exponential power delay profile defined in IEEE 802.11b \cite{cho2010mimo}, which is denoted as the exponential (EXP) channel. Therefore, our proposed AI receiver,  FC-DNN and ComNet are trained offline under the EXP channel model to perform the OTA test using the same SNR and by a specific transmitter antenna gain.

%The conventional LS+OFDM method is used as the baseline. As can be seen from Table. \ref{tableBER}, in Scenario 1, our AI receiver method achieves similar BER performance with LS+OFDM and slightly better than the other two AI-aided OFDM receivers, and the FC-DNN receiver slightly outperforms the ComNet receiver. The main reason is that the  OTA scenarios have limited transmission distances and obstacles, which lead to simple channel realizations. In Scenario 2, we establish a more complexed channel environment that increases the moving scatterers. The test results show that the LS+OFDM, FC-DNN and ComNet receivers have similar poor performance in the OTA test. Our AI-aided OFDM receiver shows its advantage in the abovementioned complex real channels since they are designed to deal with nonlinear and complex channel conditions by using nonlinear functions.

The conventional LS+OFDM method is used as the baseline. In Scenario 1 in Table \ref{tableBER}, our AI receiver method achieves similar BER performance with LS+OFDM and slightly outperforms the other two AI-aided OFDM receivers, whereas the FC-DNN receiver is slightly better than the ComNet receiver. The main reason for such outcomes is that this OTA scenario has limited transmission distances and obstacles, which leads to simple channel realizations. In Scenario 2, we establish a more complex channel environment by increasing the moving scatters. Test results show that the LS+OFDM, FC-DNN, and ComNet receivers have poor performance in the OTA test. Our AI-aided OFDM receiver shows its advantage in the abovementioned complex real channels because it is designed to deal with nonlinear and complex channel conditions by using nonlinear functions.
	\begin{table}[!h]
		\centering	
		%\footnotesize
	\caption{BER of AI receivers for OTA test}
		\begin{tabular}{>{\sf }ccccc}    %
			\toprule
			& LS+OFDM & FC-DNN &ComNet& OAMP-NET\\
			\midrule
			\multirow{1}{*}{Scenario 1}  & $3.9\times10^{-4}$ & $5.4\times10^{-4}$& $7.8\times10^{-4}$& $3.7\times10^{-4}$  \\
			
			%\hline
			\multirow{1}{*}{Scenario 2}    & $1.9\times10^{-2}$ & $2.1\times10^{-2}$& $1.2\times10^{-2}$&$4.6\times10^{-3}$ \\
			\bottomrule
		\end{tabular}		
		\label{tableBER}
	\end{table}

{Simulation results show that the designed AI receiver for a CP-free OFDM system is with lower complexity compared with the model-based receiver since it can detect the signal more precisely with fewer iterations.  Also, it has significant performance and approaches the low bound. Furthermore, the AI receiver reveals strong robustness towards comb pilots and changing environment, such as modulation modes and channel switching. Obviously, a CP-free OFDM system has larger spectrum efficiency and has more capacity space. Compared with the known AI receiver, our AI receiver has lowest BER according to both simulation and experiment. The FC-DNN is a structure of black box and can work well in simple channel environment. However, when the environment changes, it may not adjust to the new environment and work well. The ComNet is a data-driven DL model, where the channel estimation is based on LMMSE and the detection is based on zero forcing (ZF) algorithm. The ComNet may have the ability to adjust to the new channels so that the performance of the ComNet is better than that of the FC-DNN in scenario 2.  For our AI receiver, it is also a model-driven AI receiver where its fundamental models include LMMSE channel estimation and OAMP signal detection, to guarantee its baseline. Compared with the ComNet, the detection of OAMP-NET is more precise than ZF. From the results of simulation, we conclude that the trained AI receiver has strong robustness so it can work well when the environment changes, as the Table \ref{tableBER} shows}.

\section{Conclusion and Future Challenges}
%Both of channel estimation and signal detection are great challenges in CP removal cases for OFDM system. In this paper,  an AI-aided OFDM receiver is proposed in which the channel is estimated by CE-NET and the signal is detected by OAMP-NET. The OAMP-NET introduces some trainable parameters and reduces the difficulty of tuning parameters by removing the damping of OAMP algorithm. We demonstrate the superiority of the proposed receiver to recover bits in the CP-free OFDM system. The simulation results indicate that the proposed AI receiver not only has low complexity and great robustness, but also offers better BER performance than existing algorithms. In addition, the OTA test demonstrates the validity and flexibility for real channel environment.

Both channel estimation and signal detection are challenging for CP-free OFDM systems. In this paper, an AI-aided OFDM receiver has been developed, in which the channel is estimated by the CE-NET and the signal is detected by the OAMP-NET. The OAMP-NET introduces some trainable parameters and reduces the difficulty of tuning parameters by removing the damping of the OAMP algorithm. We demonstrate the superiority of the proposed receiver in recovering transmit data in the CP-free OFDM system. Simulation and experimental results indicate that the proposed AI receiver has low complexity, great robustness, and better BER performance than the existing algorithms.

%Although AI receiver has the potential to outperform conventional CP-free solutions, a performance gap may occur between simulation and the OTA test due to the difference between model channel and real environments. It is challenging to consider all possible effects in implementations to collect suitable training dataset and improve robustness of the AI-aided OFDM receivers during offline training phase. Online training is a promising method to address the problem. In future, we will concentrate on the online training to further improve the adaptation with different channel.

%Although the AI receiver has the potential to outperform conventional CP-free solutions, a performance gap may occur between the simulation and the OTA test because of the difference between the model channel and real environments. Considering all possible effects in implementation to collect a suitable training dataset and improve the robustness of the AI-aided OFDM receivers during the offline training phase is difficult. Online training is a promising method to address such problem. In future, we will concentrate on online training to further improve the adaptation of the proposed receiver  to different channels.

Although the AI receiver has the potential to outperform the conventional CP-free solutions, there is a performance gap between the continuous and comb pilots because of the estimated CSI is less accurate in the latter situation. Continuous method performs much better but requires numerous pilot symbols to obtain a reliable channel estimation, which reduces the achievable throughput of the system. Therefore, the performance improvement, especially the channel estimation, is crucial and confronts huge challenges in the case of comb pilots. Channel estimation quality can be improved by using the information in the unknown data symbols instead of only using the pilot sequences. In future, we will combine channel estimation and signal detection together to jointly acquire more accurate CSI and data symbols using smaller number of training pilots.
	\bibliographystyle{IEEEtran}
\end{document}